\documentclass[11pt,a4paper]{article} \usepackage{jheppub}
\usepackage{graphics,amsmath,amssymb}
\usepackage{xcolor}
\usepackage[rightcaption]{sidecap}
\sidecaptionvpos{figure}{c}
\usepackage{marginnote}
\usepackage{cancel}
\usepackage{ulem}
\usepackage{tikz}

\newcommand{\dd}{\mathrm{d}}

\title{Quarkonia formation in a holographic gravity-dilaton background describing QCD thermodynamics}

\author[a,b]{R. Z\"ollner} 
\author[a,b]{and B. K\"ampfer}

\affiliation[a]{Helmholtz-Zentrum  Dresden-Rossendorf, 01314 Dresden, Germany}
\affiliation[b]{Institut f\"ur Theoretische Physik, TU~Dresden, 01062 Dresden, Germany}

\emailAdd{kaempfer@hzdr.de}
\abstract{A holographic model of probe quarkonia is presented, where the dynamical gravity-dilaton background is adjusted to the thermodynamics of 
2 +1 flavor QCD with physical quark masses. The quarkonia action is modified to account for a systematic study
of the heavy-quark mass dependence. 
We focus on the $J/\psi$ and $\Upsilon$ spectral functions
and relate our model to heavy quarkonia formation
as a special aspect of hadron phenomenology in heavy-ion collisions at LHC.} 

\begin{document}

\maketitle

\section{Introduction} \label{Intro}

Heavy-quark flavor degrees of freedom receive currently some interest 
as valuable probes
of hot and dense strong-interaction matter produced in heavy-ion collisions at
LHC energies. The information encoded, e.g.\ in quarkonia ($c \bar c$, $b \bar b$)
observables,  supplements
penetrating electromagnetic probes and hard (jet) probes and the rich flow observables,
thus complementing each other in characterizing the dynamics 
of quarks and gluons up to the final hadronic states 
(cf.\ contributions in \cite{Proceedings:2019drx} for the state of the art). 
Since heavy quarks emerge essentially in early, hard processes, they witness
the course of a heavy-ion collision -- either as individual entities or subjects  of
dissociating and regenerating bound states 
\cite{Strickland:2019ukl,Prino:2016cni,Yao:2018sgn}. 
Accordingly, the heavy-quark
physics addresses such issues as charm ($c$, $\bar c$) and bottom ($b$, $\bar b$)
dynamics related to transport coefficients 
\cite{Rapp:2018qla,Xu:2018gux,Cao:2018ews,Brambilla:2019tpt,Song:2019cqz} 
in the rapidly evolving and highly
anisotropic ambient quark-gluon medium
\cite{Chattopadhyay:2019jqj,Bazow:2013ifa}
as well as $c \bar c$ and $b \bar b$
states as open quantum systems 
\cite{Katz:2015qja,Blaizot:2017ypk,Blaizot:2018oev,Brambilla:2017zei}. 
The rich body of experimental data from LHC,
and also from RHIC, enabled a tremendous refinement of our 
understanding heavy-quark dynamics.
For a recent survey on the quarkonium physics we refer the interested reader to \cite{Rothkopf:2019ipj}.  

The yields of various hadron species, light nuclei and anti-nuclei 
-- even such ones which are only very loosely bound 
\cite{Braun-Munzinger:2018hat} -- 
emerging from heavy-ion collisions at LHC energies 
are described by the thermo-statistical hadronization model \cite{Andronic:2017pug}
with high accuracy. These yields span an interval of nine
orders of magnitude. 
The final hadrons and nuclear clusters are described by two parameters: 
the freeze-out temperature $T_{fo} =155$~MeV and a freeze-out volume depending on the system size or centrality of the collision. Due to the
near-perfect matter-antimatter symmetry at top LHC energies the 
baryo-chemical potential $\mu_B$ is exceedingly small, $\mu_B/ T_{fo} \ll 1$. 
It is argued in \cite{Andronic:2017pug} that the freeze-out of color-neutral
objects happens just in the demarcation region of hadron matter to quark-gluon plasma,
i.e. confined vs. deconfined strong-interaction matter. 
In fact, lattice QCD results
report a pseudo-critical temperature of 
$T_c = (156 \pm 1.5)$~MeV \cite{Bazavov:2018mes} -- a value agreeing with the disappearance of the chiral condensates and the maximum of some susceptibilities. 
The key is the adjustment of physical quark masses and the use of 2+1 
flavors \cite{Borsanyi:2013bia,Bazavov:2014pvz}, 
in short QCD$_{2+1}$(phys). Details of the (may be accidental)
coincidence of deconfinement and chiral symmetry restoration are matter of debate 
\cite{Suganuma:2017syi}, 
as also the formation of color-neutral objects out of the cooling quark-gluon plasma at $T_c$.  For instance, Reference \cite{Bellwied:2018tkc} advocates
flavor-dependent freeze-out temperatures. 
Note that at $T_c$ no phase transition happens, rather the thermodynamics
is characterized by a cross-over accompanied by a pronounced dip in the sound
velocity. 

Among the tools for describing hadrons as composite strong-interaction
systems is holography. Anchored in the famous AdS/CFT correspondence,
holographic bottom-up approaches have facilitated a successful description of
mass spectra, coupling strengths/decay constants etc.\ of various hadron species.
While the direct link to QCD by a holographic QCD-dual or rigorous top-down
formulations are still missing, one has to restrict the accessible observables 
to explore certain frameworks and scenarios. We consider here a framework
which merges (i) QCD$_{2+1}$(phys) thermodynamics described 
by a dynamical holographic gravity-dilaton
background and (ii) holographic probe quarkonia.
We envisage a scenario which embodies QCD thermodynamics
of QCD$_{2+1}$(phys) and the emergence of hadron states 
at $T_c$  at the same time.
One motivation of our work is the exploration of a holographic model which is in agreement with the above hadron 
phenomenology in heavy-ion collisions at LHC energies. Early holographic attempts
\cite{Colangelo:2012jy,Colangelo:2009ra,Colangelo:2008us} to hadrons at non-zero temperatures faced the problem of meson melting
at temperatures significantly below the deconfinement temperature $T_c$.
Several proposals have been made \cite{Zollner:2016cgc,Zollner:2017fkm,Zollner:2017ggh,Braga:2015lck} to find
rescue avenues which accommodate hadrons at and below $T_c$.
Otherwise, a series of holographic models of hadron melting
without reference to QCD thermodynamics, 
 e.g.\
\cite{Braga:2015lck,Braga:2016wkm,Fujita:2009wc,Fujita:2009ca,Grigoryan:2010pj,Braga:2017bml,Andreev:2019hrk,Vega:2018dgk,Mamani:2018uxf,Dudal:2014jfa},
finds quarkonia states well above, at and below $T_c$
in agreement with lattice QCD results
\cite{Bazavov:2014cta,Kim:2018yhk,Ding:2019kva,Larsen:2019zqv}. 
It is therefore tempting to
account for the proper QCD-related background. 

In the temperature region $T \approx{\cal O} (T_c)$, the impact of charm and bottom
degrees of freedom on the quark-gluon--hadron thermodynamics is minor \cite{Borsanyi:2016ksw}. 
Thus, we consider quarkonia as test particles. We follow 
\cite{Gubser:2008ny,Finazzo:2014cna,Finazzo:2013efa,Zollner:2018uep} 
and model  the holographic background 
by a gravity-dilaton set-up, i.e.\ without adding further fundamental degrees
of freedom (as done, e.g.\ in \cite{Bartz:2018nzn,Bartz:2016ufc,Bartz:2014oba})
to the dilaton, which was originally related solely to gluon degrees
of freedom \cite{Gursoy:2010fj}. That is, the dilaton potential is adjusted to QCD$_{2+1}$(phys) lattice data. 
Our emphasis is on the formation of quarkonia in a cooling
strong-interaction environment. Thereby, the quarkonia properties
are described by spectral functions. 
We restrict ourselves to equilibrium and leave non-equilibrium effects,
e.g.\ \cite{Bellantuono:2017msk,Yao:2017fuc}, for future work.

Our paper is organized as follows. In section \ref{sect:quarkonia}, the dynamics of the probe quarkonia is formulated and the coupling to the thermodynamics-related background
is explained.  (The recollection of the gravity-dilaton dynamics and the consideration of special features are relegated to appendix \ref{App:B}.) 
Numerical solutions in the charm ($J/\psi$) and bottom ($\Upsilon$) sectors
w.r.t.\ quarkonium spectral functions and the quarkonium formation systematic 
are dealt with in section \ref{sect:two_param}. The tested two-parameter
Schr\"odinger potential facilitates bottomonium formation as rapid squeezing
of the spectral function towards a narrow quasi-particle state in a small
temperature interval around $T_c$. An analogous behavior is accomplished for
charmonium by a three-parameter potential considered in section
\ref{sect:three_param}. The squeezing of the charmonium spectral function extends
over a somewhat longer temperature interval and requires a particular
parameter setting. We summarize in section \ref{sect:summary}. 
 
\section{Quarkonia as probe vector mesons}\label{sect:quarkonia}

The action of quarkonia as probe vector mesons in string frame is 
\begin{equation} \label{eq:1}
S_m^V = \frac{1}{k_V} \int \dd^4x \, \dd z \sqrt{g_5} e^{-\phi} \, G_m(\phi) \,F^2 ,
\end{equation}
where the function $G_m(\phi)$ carries the flavor 
(or heavy-quark mass, labeled by $m$)
dependence
and $F^2$ is the field strength tensor squared of a $U(1)$ gauge field ${\cal A}$
in 5D asymptotic anti-de Sitter (AdS) space time, 
with or without black hole (BH),
with the bulk coordinate $z$ 
and metric fundamental determinant $g_5$; 
$\phi$ is the scalar dilatonic field with zero mass dimension.  
The gauge field ${\cal A}$ in the bulk is sourced by a current operator
of the structure $\bar Q \gamma_\mu Q$ at the boundary, where
$Q$ stands for the heavy quark field operator. 
The structure of (\ref{eq:1}) is that of a field-dependent gauge kinetic term,
familiar, e.g., from realizations of a localization mechanism in 
brane world scenarios
\cite{Chumbes:2011zt,Eto:2019weg,Arai:2017lfv}.
In holographic Einstein-Maxwell-dilaton models (cf.~\cite{DeWolfe:2010he}), 
often employed in including a conserved charge density (e.g.~\cite{Rougemont:2015wca,Knaute:2017opk}),
such a term refers to the gauge coupling.

The action (\ref{eq:1}) with $G_m = 1$, originally put forward 
in the soft-wall (SW) model for light-quark mesons \cite{Karch:2006pv}, 
is also used for describing heavy-quark vector mesons 
\cite{Braga:2016wkm,Fujita:2009wc,Fujita:2009ca},
e.g.\
charmonium \cite{Grigoryan:2010pj,Braga:2017bml} 
or bottomonium \cite{Braga:2018zlu}. 
As emphasized, e.g.\ in \cite{Grigoryan:2010pj}, 
the holographic background encoded in $g_5$ and $\phi$ must be chosen
differently to imprint the different mass scales, since (\ref{eq:1}) with $G_m =1$
as such would be flavor blind. 
Clearly, the combination $\exp\{- \phi\} G_m(\phi)$ in (\ref{eq:1})
with flavor dependent function $G_m(\phi)$
is nothing but introducing effectively a flavor dependent dilaton profile
$\phi_m = \phi - \log G_m$,
while keeping the thermodynamics-steered hadron-universal dilaton $\phi$.
In fact, many  authors use the form 
$S_m^V = \frac{1}{k_V} \int \dd^4x \, \dd z \sqrt{g_5} e^{-\phi_m} \,F^2$
to study the vector meson melting by employing different parameterizations
of $\phi_m$ to account for different flavor sectors. Here, we emphasize
the use of a unique gravity-dilaton background for all flavors 
and include the quark mass (or flavor) dependence solely in $G_m$.

Our procedure to determine $G_m$ is based on the import of information from the hadron sector at $T = 0$.  
The action  (\ref{eq:1}) leads via the gauges ${\cal A}_z  = 0$ and 
$\partial^\mu {\cal A}_\mu = 0$ and the ansatz
${\cal A}_\mu = \epsilon_\mu \, \varphi (z) \, \exp\{ i p_\nu x^\nu \}$
with $\mu, \nu = 0, \cdots,3$, which uniformly separates the $z$ dependence
of the gauge field by the bulk-to-boundary propagator $\varphi$ for all components of ${\cal A}$, and the constant polarization vector $\epsilon_\mu$
to the equation of motion 
\begin{equation} \label{eq:EoM}
\varphi'' +
\left[\frac12 A' + (\partial_\phi \log G_m -1 ) \phi' + (\log f)' \right] \varphi' +
\frac{p^\mu p_\mu}{f^2} \varphi = 0,
\end{equation}
where $A(z, z_H)$ is the warp factor and $f(z, z_H)$ denotes the
blackening function in the AdS + BH metric with horizon at $z_H$,
\begin{equation} \label{eq:3}
\dd s^2 = \exp\{ A(z, z_H)\} 
\left[f(z, z_H) \, \dd t^2 - \dd \vec x^2 - \frac{\dd z^2}{f(z, z_H)} \right] ,
\end{equation}
and a prime denotes the derivative w.r.t.\ the bulk coordinate $z$.
Both, $A(z, z_H)$ and $f(z, z_H)$, are solutions of Einstein's equation with a dilatonic potential adjusted to QCD thermodynamics with physical quark masses
in the temperature range 100~MeV $< T <$ 400~MeV (cf.\ appendix A
in \cite{Zollner:2020cxb} and appendix \ref{App:B} for details); 
also $\phi(z, z_H)$ is determined dynamically and is consistent with the
metric coefficients via field equations.

By the transformation 
$\psi (\xi) = \varphi(z (\xi)) \, \exp\{ \frac 12 \int_0^\xi \dd z  \, {\cal S}_T (\xi) \}$
one gets the form
of a one-dimensional Schr\"odinger equation with 
the tortoise coordinate $\xi$
\begin{equation} \label{eq:6}
\left[\partial_\xi^2 - (U_T(z(\xi)) - m_n^2) \right] \psi_n (\xi)= 0, 
\quad n = 0, 1, 2,\cdots,
\end{equation} 
where one has to employ $z(\xi)$ from solving $\partial _\xi = (1/f) \partial_z$.
The Schr\"odinger equivalent potential is  
\begin{equation} \label{eq:7}
U_T \equiv \left( \frac12 {\cal S}_T' + \frac14 {\cal S}_T^2 \right) f^2 
+ \frac12 {\cal S}_T f f' 
\end{equation}
as a function of $\xi(z)$ with
\begin{equation} \label{eq:G}
{\cal S}_T \equiv  \frac12 A' - \phi' + \partial_z \log G_m (\phi(z)).
\end{equation}

At $T = 0$ (label ``0"), $f = 1$ and $\xi \to z$ and $U_T \to U_0$ with
\begin{eqnarray} 
U_0 (z) & \equiv & \frac12 {\cal S}_0' + \frac14 {\cal S}_0^2 , \label{eq:8}\\
{\cal S}_0& \equiv &  \frac12 A_0' (z) - \phi_0' (z) + 
\partial_z \log G_m (\phi_0(z)) , \label{eq:9}
\end{eqnarray}
and (\ref{eq:6}) becomes 
\begin{equation} \label{eq:S}
\left[ \partial_z^2 + (U_0 (z) - m_n^2) \right] \psi_n = 0
\end{equation}
with normalizable solutions $\psi_n$ and discrete states with
masses squared $m_n^2 = p^\mu p_\mu$, $n = 0, 1, 2, \cdots$
for quarkonia at rest.
That is, at $T = 0$ one has to deal with a suitable Schr\"odinger equivalent 
potential $U_0 (z)$ to generate the desired spectrum $m_n$. 
In such a way, the needed hadron physics information at $T = 0$ is imported
by parameterizing $U_0$ in suitable manner
(see sections \ref{sect:two_param} and \ref{sect:three_param}).

The next step is solving (\ref{eq:8}) to obtain ${\cal S}_0 (z)$
and, with (\ref{eq:9}), then $G_m(\phi)$ with $G_m (0) = 1$.
This needs $A_0'(z)$ and $\phi_0'(z)$, which follow from the thermodynamics sector 
(see Appendix \ref{App:B})
via $A_0 = A(z) = \lim_{z_H \to \infty} A(z, z_H)$ and 
$\phi_0 = \phi(z) = \lim_{z_H \to \infty} \phi(z, z_H)$.
One has to suppose that these limits are meaningful.
The limited information from lattice QCD thermodynamics 
w.r.t.\ the finite temperature range and the data accuracy may pose here a
problem. Ignoring such a potential obstacle we use then 
$G_m(\phi) = G_m(z(\phi_0))$ as universal (i.e.\ temperature independent)
function.

The equation of motion (\ref{eq:EoM}) of $\varphi$
can also be employed to compute quarkonia spectral functions, 
cf.~\cite{Colangelo:2012jy,Fujita:2009wc,Fujita:2009ca,Grigoryan:2010pj,Hohler:2013vca}.
For $\omega^2  = p^{\mu} p_{\mu} > 0$ fixed, the asymptotic
boundary behavior facilitates two linearly
independent solutions by considering the leading order terms
on both sides of the interval $[0,z_H]$.
(i) For $z \to 0$, one has, due to the AdS asymptotics at the boundary,
the general solution 
\begin{equation}
\varphi(z \to 0) \to A(\omega) \varphi_1 + B(\omega) \varphi_2
\end{equation}
with two $\omega$-dependent complex constants $A$ and $B$, 
and $\varphi_1(z \to 0) \to 1$ and  $\varphi_2(z \to 0) \to (z/z_H)^2$.
(ii) Near the horizon, $z \to z_H$, the asymptotic behavior of solutions of
(\ref{eq:EoM}) is
steered by the poles of $1/f$ and $1/f^2$. 
The two linearly independent solutions are
$\varphi_{\pm} (z \to z_H) \to (1 - \frac{z}{z_H})^{\pm i \omega /|f'(z_H)|}$,
where $\varphi_{\pm}$ represent out-going and in-falling solutions,
respectively.
Then, the general near-horizon solution is given by
\begin{equation}
\varphi(z \to z_H) \to C(\omega) \varphi_+ + D(\omega) \varphi_- ,
\end{equation}
again with complex constants $C$ and $D$ which depend on $\omega$. 
The obvious and commonly used side conditions for 
the bulk-to-boundary propagator are $\varphi(0)=1$, 
which means $A(\omega)=1$, and 
$\varphi(z \to z_H) = \varphi_- (z \to z_H)$
(purely in-falling solution at the black hole horizon), 
yielding $C(\omega)=0$. 
Due to the bilinear mapping $(A,B) \mapsto (C,D)$,  
the value of $B$ for getting the desired
in-falling solution can be easily determined by solving (\ref{eq:EoM})
twice, once with $A=1$, $B=0$ and once with $A=0$, $B=1$ 
and comparing the result with $\varphi_-$ to dig out the proper
coefficients.

In more detail, the first integration starts with some sufficiently small
$\varepsilon$, thus  setting the near-boundary initial conditions for $\varphi$
equal to $\varphi_1$, i.e.
$\varphi(z_H\varepsilon)=1$ and
$\varphi'(z_H\varepsilon)=0$. 
Near the horizon, at $z = z_H (1 - \varepsilon)$,
the obtained value $y_1$ of this solution and 
the value $y_1'$ of its derivative can be written as
superposition of $\varphi_+$ and $\varphi_-$:
  \begin{eqnarray}
   y_1&=& C_1 \varphi_+(z_H(1-\varepsilon)) +
D_1\varphi_-(z_H(1-\varepsilon)) , \\
   y_1'&=& C_1 \varphi_+'(z_H(1-\varepsilon)) +
D_1\varphi_-'(z_H(1-\varepsilon)).
  \end{eqnarray}
The constants $C_1$, $D_1$ are determined by solving this linear system. 
The second integration works analogously, now based on $\varphi_2$
for near-boundary initial values of $\varphi$ and its derivative,
i.e. $\varphi(z_H\varepsilon)=\varepsilon^2$ and
$\varphi'(z_H\varepsilon)=2 \varepsilon /z_H$, thus yielding another
solution, which is decomposed as 
$y_2= C_2 \varphi_+(z_H(1-\varepsilon)) + D_2\varphi_-(z_H(1-\varepsilon))$.
This, together with $y_2'$, determines $C_2$ and $D_2$. 
A straightforward calculation using the above mentioned linearity shows that the
particular value $B = - C_1/C_2$ eliminates the out-going part of the
general solution near the horizon \cite{Teaney:2006nc}.

Then, the corresponding retarded Green function $\mathcal{G}^{\rm R}$ of
the dual current operator $\bar Q \gamma_\mu Q$, defined within the framework of the holographic dictionary 
via a generating functional by
$\mathcal{G}^{\rm R} = \frac{\delta^2}{\delta {\cal A}^{0 \, \mu} (-\omega)
\delta {\cal A}^0_\mu (\omega)}
\langle \exp\{i \int \dd^4 x \, {\cal A}^0_\nu \bar Q \gamma^\nu Q \} \rangle $,
is given by (cf.\ \cite{Teaney:2006nc})
\begin{equation}
\mathcal{G}^{\rm R} (\omega) =
\frac{\delta^2 S_m^{V,\, \mbox{on-shell}}}{\delta \mathcal{A}^{0 \, \mu} (-
\omega) \, \delta \mathcal{A}^0_\mu (\omega)} 
= k \lim \limits_{z \to 0} \frac1z
\varphi^* (z) \varphi' (z)
= \frac{2 k}{z_H^2} B(\omega)
\end{equation}
with $k=\frac{N_c}{24\pi^2}$ and
$\mathcal{A}^0_\mu \equiv \epsilon_{\mu} \exp\{ip_{\nu}x^{\nu}\}$
for $\mu \in \{1, 2, 3 \}$.
The quantity $S_m^{V,\, \mbox{on-shell}}$ denotes here the
action (\ref{eq:1}) with the solution $\varphi$ from (\ref{eq:EoM}).
Finally, the spectral function $\rho$ follows from
$\rho(\omega) =  
\mathrm{Im} \, \mathcal{G}^{\rm R} (\omega)= \frac{2 k}{z_H^2} \mathrm{Im} \, B (\omega)$.
It has the dimension of energy squared.

\section{Two-parameter potential -- bottomonium formation} 
\label{sect:two_param}
 
Our setting does not explicitly refer to a certain quark mass $m$. Instead,
an ansatz $U_0(z; \vec p \, )$ with parameter $n$-tuple $\{\vec p \, \}$
is used such to catch a certain quarkonium mass spectrum.
Insofar, $m$ is to be considered as cumulative label highlighting the dependence
of $G_m$ on a parameter set $\{\vec p \, \}$
which originally enters $U_0$ and which is to be adjusted to
charmonium and bottomonium masses.

\begin{figure}
\centering
\resizebox{0.55\columnwidth}{!}{
\includegraphics{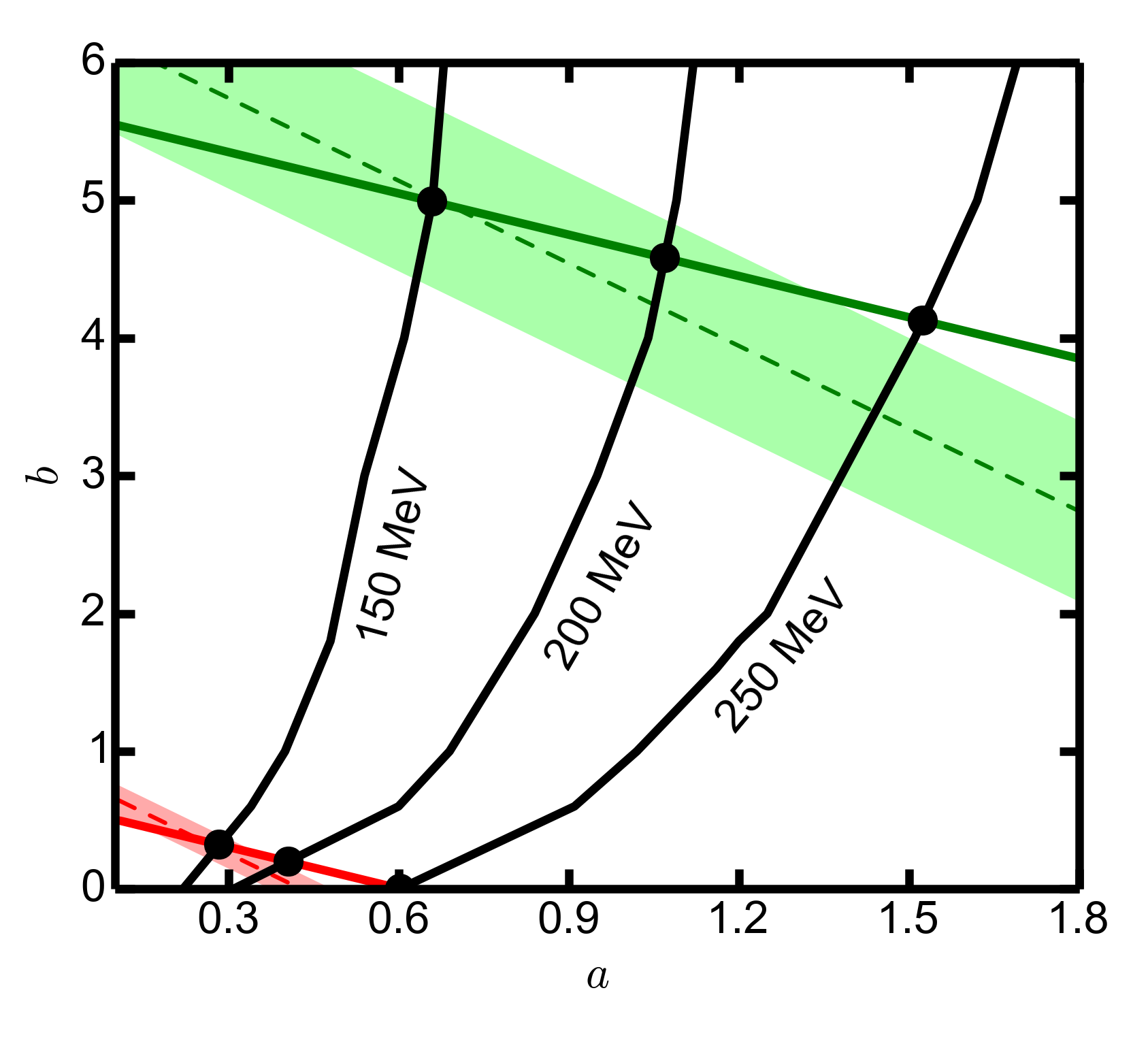}
\put(-340,275){{\huge {\color{green}$\Upsilon$}}}
\put(-355,95){{\huge {\color{red}$J/\psi$}}}
}
\caption{Constant ground state masses (fat solid lines) 
and the respective first excitations
(dashed lines; a $\pm 5$\% corridor is depicted by colored bands)
at $T = 0$ over the dimensionless $a$-$b$ plane of the potential (\ref{eq:10}).
Color code: green - $\Upsilon$, red - $J/\psi$.
The bullets mark selected parameters for the spectral functions
exhibited in figure~\ref{fig:A4}.
The black curves exhibit the loci at which the peaks of the spectral functions
completely disappear, i.e.\ they represent the contours of melting temperatures
$T_{melt}^{{\rm g.s.}} (a, b) = 150$, 200 and 250~MeV. 
That is, for a given point $(a, b)$ in the parameter space, the spectral function
in the energy range of the ground state displays a peak only for 
$T < T_{melt}^{{\rm g.s.}}$. 
\label{fig:A1}
}
\end{figure}

As a transparent model we select the two-parameter potential 
\cite{Fujita:2009wc,Grigoryan:2010pj}
\begin{equation} \label{eq:10}
U_0 (z) = \frac34 \left( \frac{1}{z} \right)^2 +  \hat a^2 z^2
+ 4 \hat b
\end{equation} 
which is known to deliver via (\ref{eq:S}) the normalizable functions $\psi_n$ with discrete eigenvalues
\begin{equation} \label{eq:11}
m_n^2 = 4 (\hat a + \hat b + n \, \hat a ), \quad n= 0, 1, 2, \cdots . 
\end{equation}
The potential (\ref{eq:10}) is a slight modification of the SW model
\cite{Karch:2006pv} with 
$3/ (4 z^2)$ stemming from the near-boundary warp factor $A(z)$ and the 
term $\propto z^2$
emerging originally from a quadratic dilaton profile ansatz.  
Note the Regge type excitation spectrum $m_n^2 = m_0^2 +  n \hat a$ with
intercept and slope to be steered by two independent parameters 
$\hat a$ and $\hat b$.  We choose these parameters as follows.
The mass $m_0$ determines the ground state (g.s.)  ``trajectory" 
in the $a$-$b$ plane, $\hat b = \frac14 m_0^2 - \hat a$,
and $m_1$ determines the first excitation (1st) ``trajectory"
 by $\hat b = \frac14 m_1^2 - 2 \hat a$.
Using the PDG values of $J/\psi, \, \psi'$ and $\Upsilon (1S, \,2S)$ adjusts the 
 ``trajectories" as solid and dashed lines in figure~\ref{fig:A1}, 
where we employ the
scale setting via $\hat a = a / L^2$ and $\hat b = b / L^2$ with
$L^{-1} = 1.99$ GeV, which is related to the QCD thermodynamics
sector (see appendix A in \cite{Zollner:2020cxb}). 
Allowing for a 10\% variation of $m_1$ one arrives at the colored
bands in figure~\ref{fig:A1}. 
By such a parameter choice one puts emphasis
on the quarkonia g.s.\ masses as representatives of the heavy quark masses
and less emphasis on the level spacing of excitations and ignores other possible
constraints.

As we shall demonstrate below, the ansatz (\ref{eq:10}) 
has several drawbacks and,
therefore, is to be considered as an illustrative example.
For instance, the sequence of radial  $\Upsilon$ excitations in nature does not form a strictly linear Regge trajectory \cite{Ebert:2011jc}.
This prevents an unambiguous mapping of $m_{0, 1} \mapsto (a, b)$.
While the radial excitations of $J/\psi$ follow quite accurately a linear Regge trajectory in nature \cite{Ebert:2011jc},
the request of accommodating further $J/\psi$ properties in $U_0$ calls also for
modifying  (\ref{eq:10}), cf.\ \cite{Grigoryan:2010pj,Hohler:2013vca}. 
Despite the mentioned deficits,
the appeal of (\ref{eq:10}, \ref{eq:11}) is nevertheless the simply invertible relation $m_n^2 (a, b)$
yielding $a(m_{0,1})$ and $b(m_{0,1})$. Since we are going to
study the systematic, we keep the primary parameters $a$ and $b$ in what follows. 
Instead of discussing results at isolated points in parameter space
referring to $J/\psi$ and $\Upsilon$ ground states $m_0$ and first
excited states $m_1$, we consider the systematic over the $a$-$b$ plane. 

\begin{figure}[t]
\center

\resizebox{0.99\columnwidth}{!}{%
\includegraphics{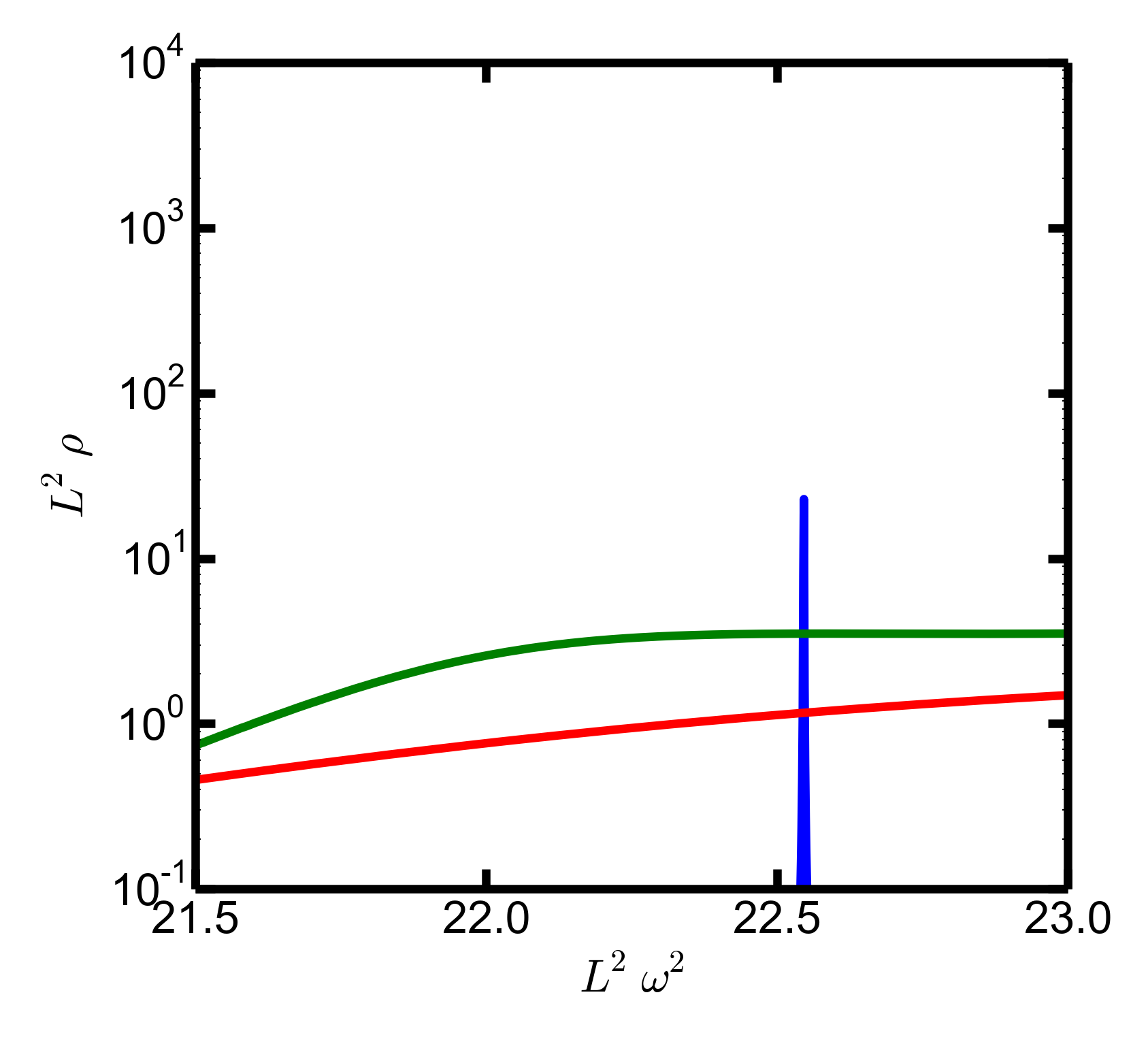}
\includegraphics{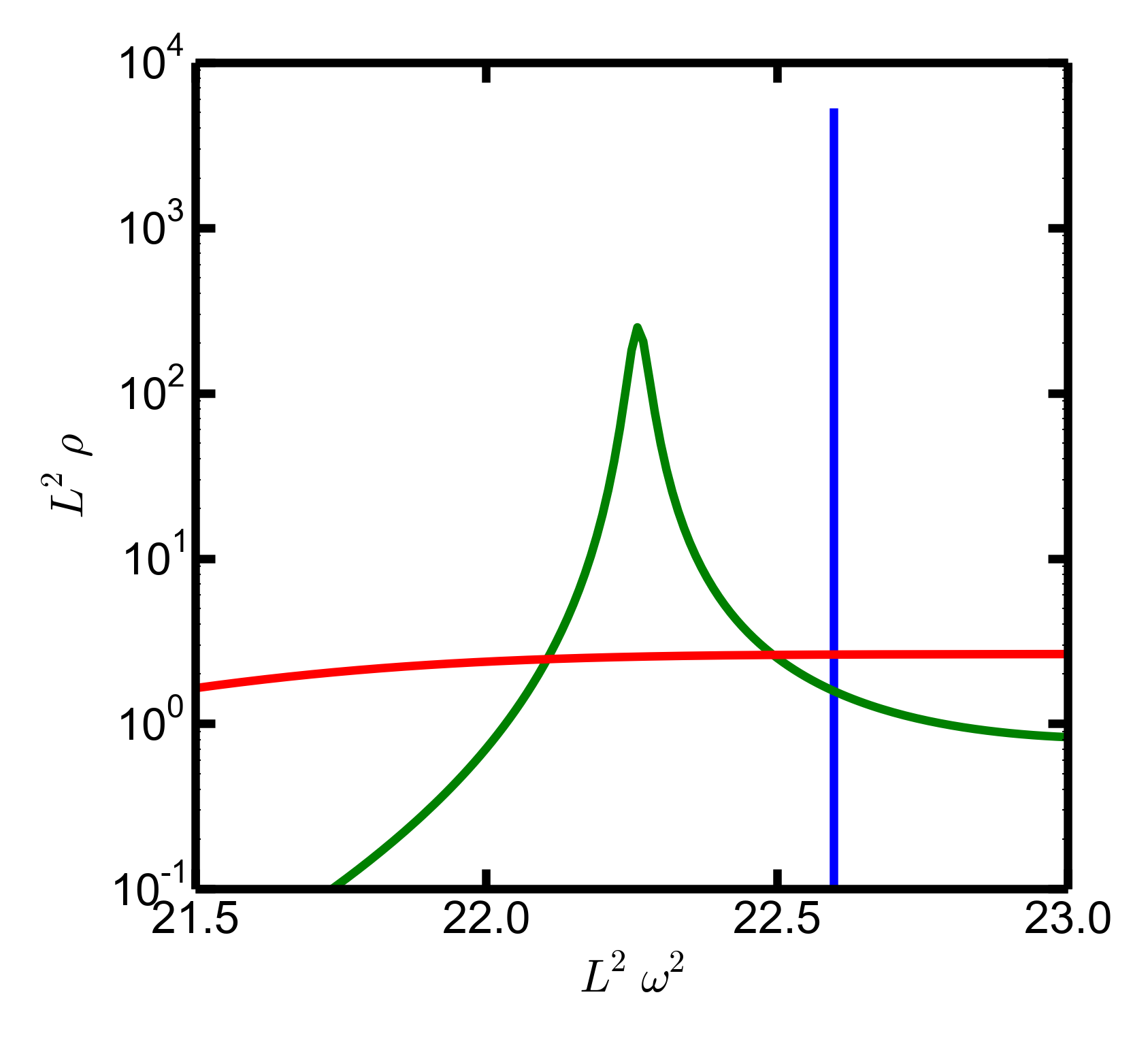}
\includegraphics{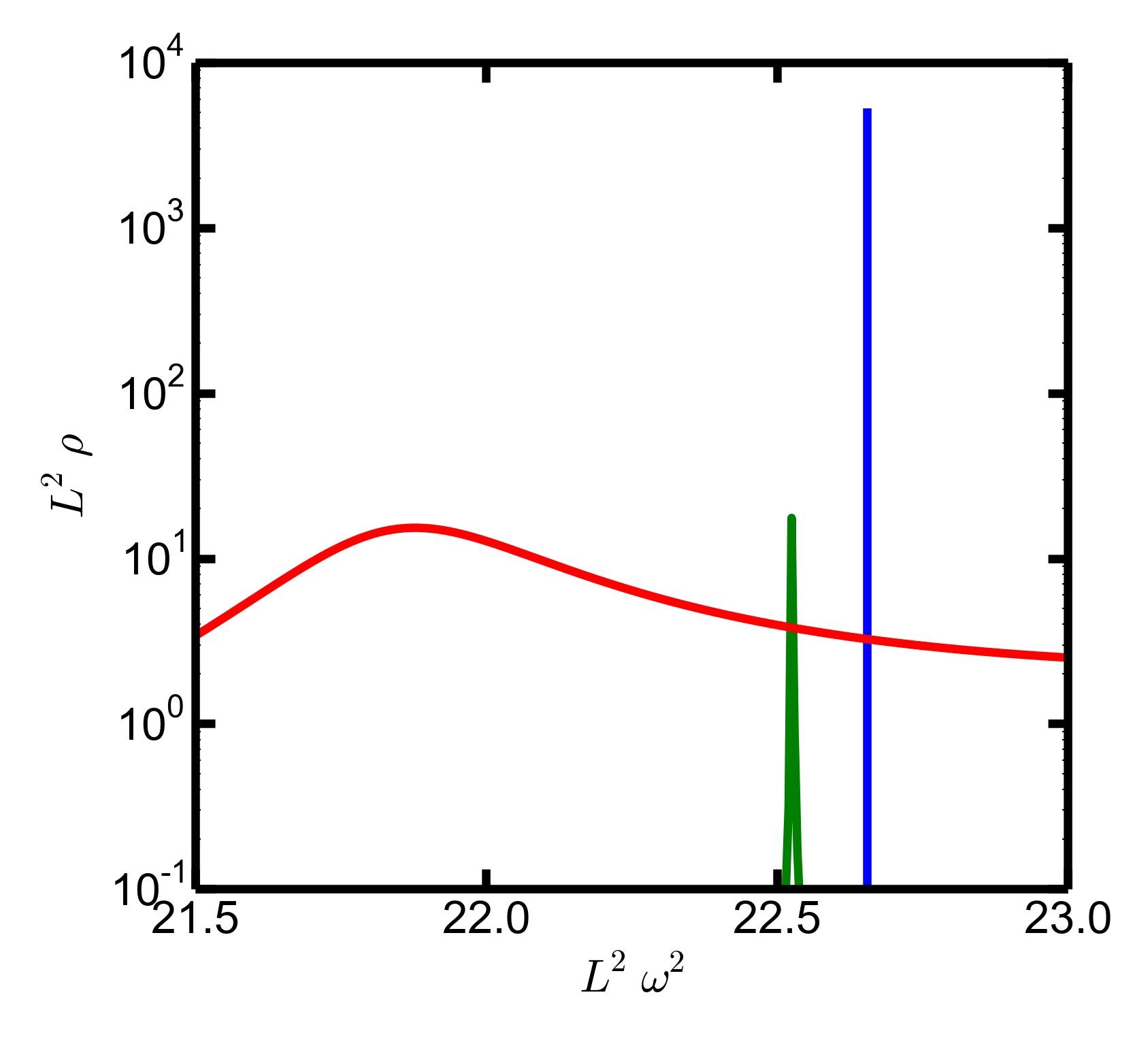}

\put(-1121,277){{\Huge $\Upsilon: m_0$, (0.6595, 4.99)}}
\put(-722,277){{\Huge $\Upsilon: m_0$, (1.069, 4.58 )}}
\put(-288,277){{\Huge $\Upsilon: m_0$, (1.5238,  4. 1257)}}
}

\resizebox{0.99\columnwidth}{!}{%
\includegraphics{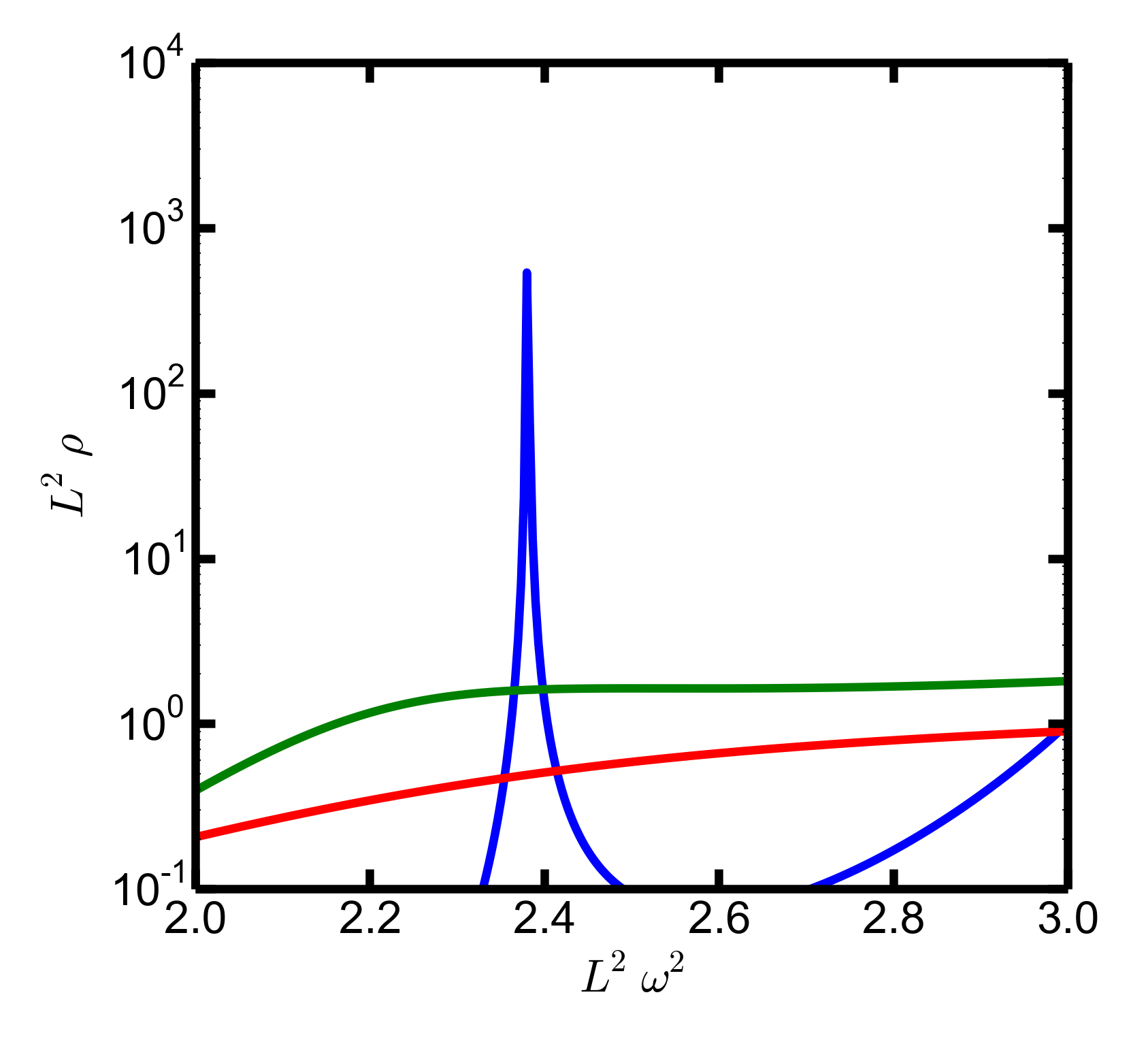}
\includegraphics{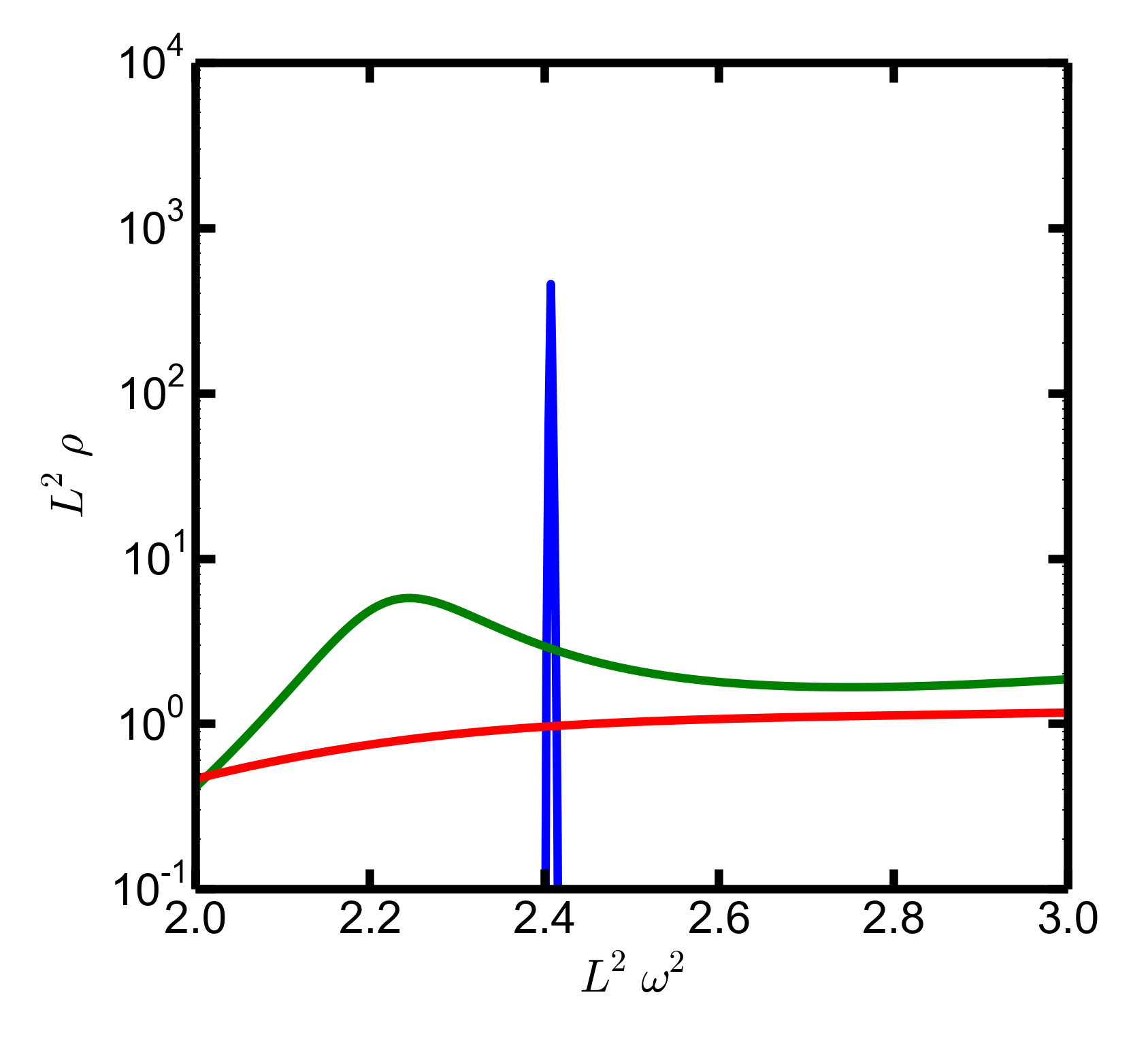}
\includegraphics{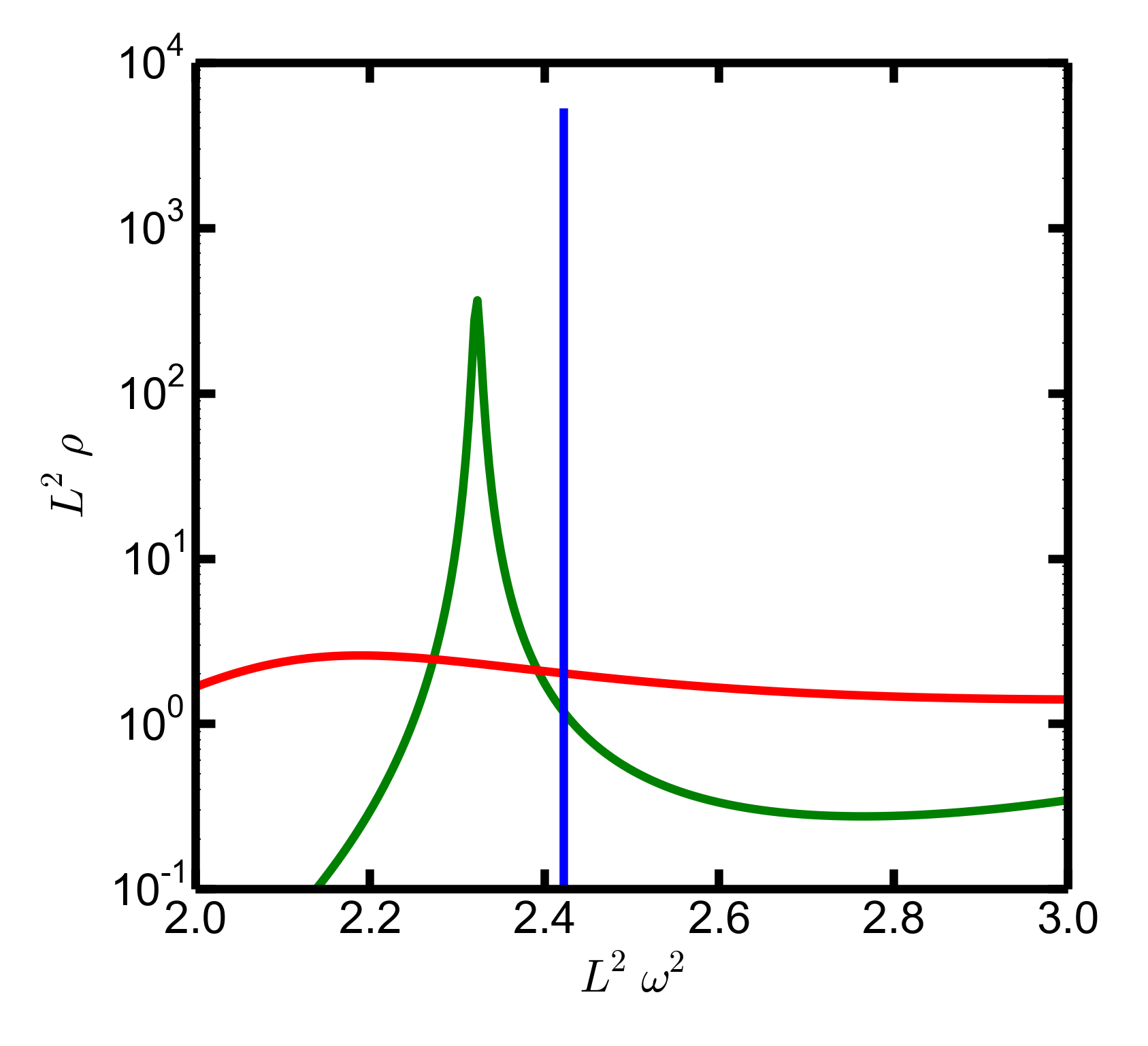}
\put(-1121,277){{\Huge $J/\psi: m_0$, (0.2842, 0.3212)}}
\put(-722,277){{\Huge $J/\psi: m_0$, \, (0.4056, 0.1999)}}
\put(-311,277){{\Huge $J/\psi: m_0$, (0.6027, 0.0027)}}
}
\caption{Spectral functions $L^2 \rho (\omega, T)$
of  $\Upsilon$ (top row) and $J/\psi$ (bottom row) in the g.s.\ energy region
at $T = 100$ (blue), 150 (green), and 200~MeV (red).
For parameters $(a, b)$ on the g.s.\ trajectories,
i.e.\ for PDG values of $m_0$.
The values of $(a, b)$ are given in the legends;
these positions are marked in figure~\ref{fig:A1} by bullets. 
\label{fig:A4}
}
\end{figure}

The black curves in figure~\ref{fig:A1} exhibit the contours
$T_{melt}^{{\rm g.s.}} (a, b) = 150$, 200 and 250~MeV.\footnote{
An analog figure in \cite{Zollner:2020cxb} exhibits the contour plot of 
the dissociation temperature $T_{dis} (a, b)$
which has been determined by the disappearance of normalizable solutions
of the Schr\"odinger equation (\ref{eq:6}) in the interval
$z = [\epsilon_0, \tilde z_H]$ with boundary conditions
$\psi_n (z = \epsilon_0) \propto \epsilon_0^2$ and 
$\psi_n (z = \tilde z_H) = 0$. $\tilde z_H = z_H ( 1- \epsilon_H)$ with
$\epsilon_H = 10^{-2}$ sets a convenient cut-off which suppresses
the highly oscillating solutions towards the horizon at $z_H$. In contrast,
the limit $\epsilon_0 \to 0$ is well defined. We find in general
$T_{melt} (a,b) > T_{dis} (a,b)$.}
The melting temperature 
$T_{melt}^{{\rm g.s.}}$ is determined by the disappearance of the peak of the
g.s.\ spectral function upon temperature increase. One observes
a strong parameter dependence as well, which determines the 
spectral functions, see figure~\ref{fig:A4}. Changing the parameters
$(a, b)$ deforms the potential (\ref{eq:10}) in a characteristic manner
\cite{Zollner:2020cxb}, e.g.\ going on a g.s.\ trajectory to right squeezes
the excited states to higher energies, as can be identified in figure~\ref{fig:A1},
in particular for the $\Upsilon$. Such changes affect immediately the spectral
functions.

\begin{figure}[t]
\center
\resizebox{0.99\columnwidth}{!}{%

\includegraphics{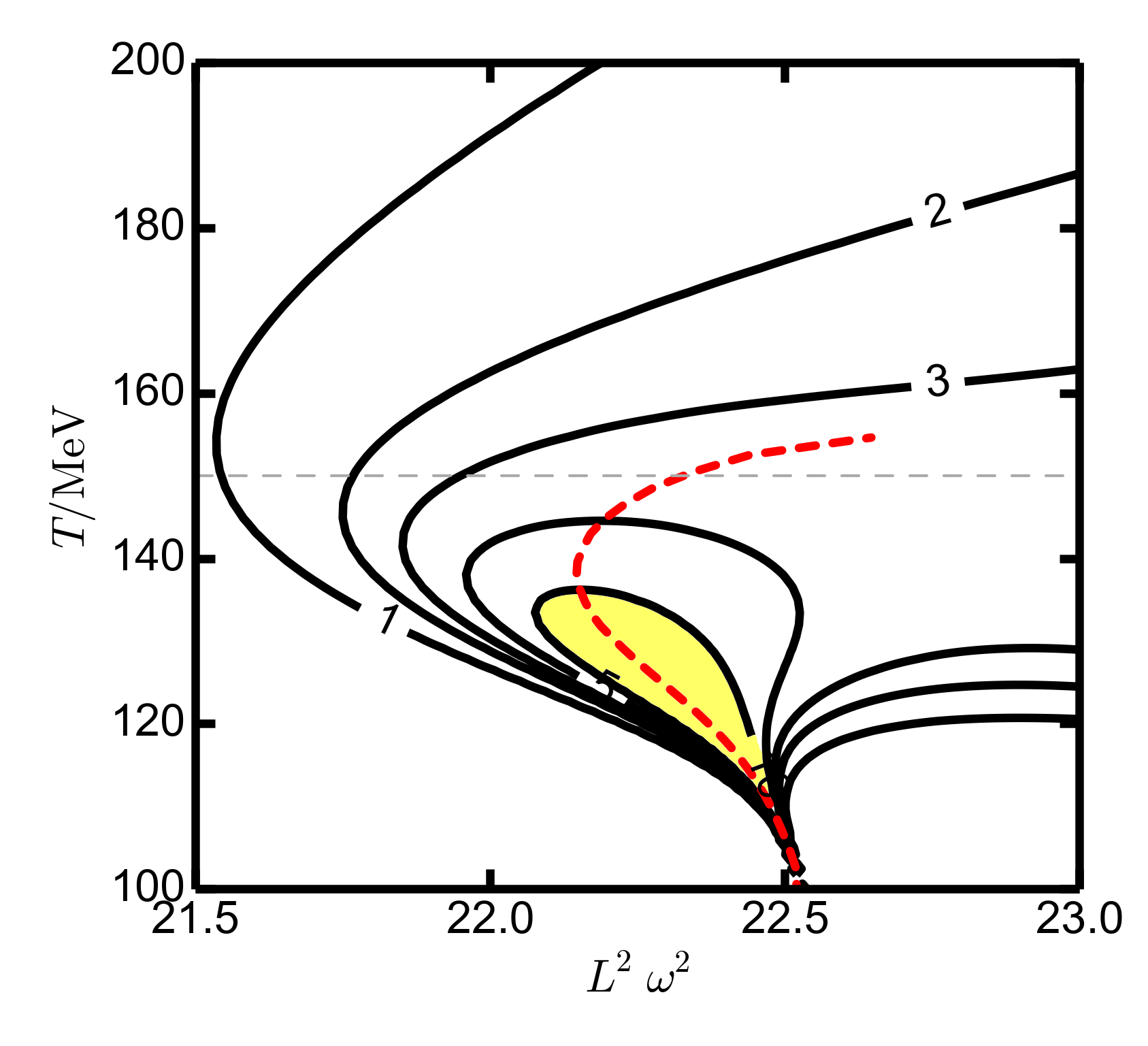}  
\includegraphics{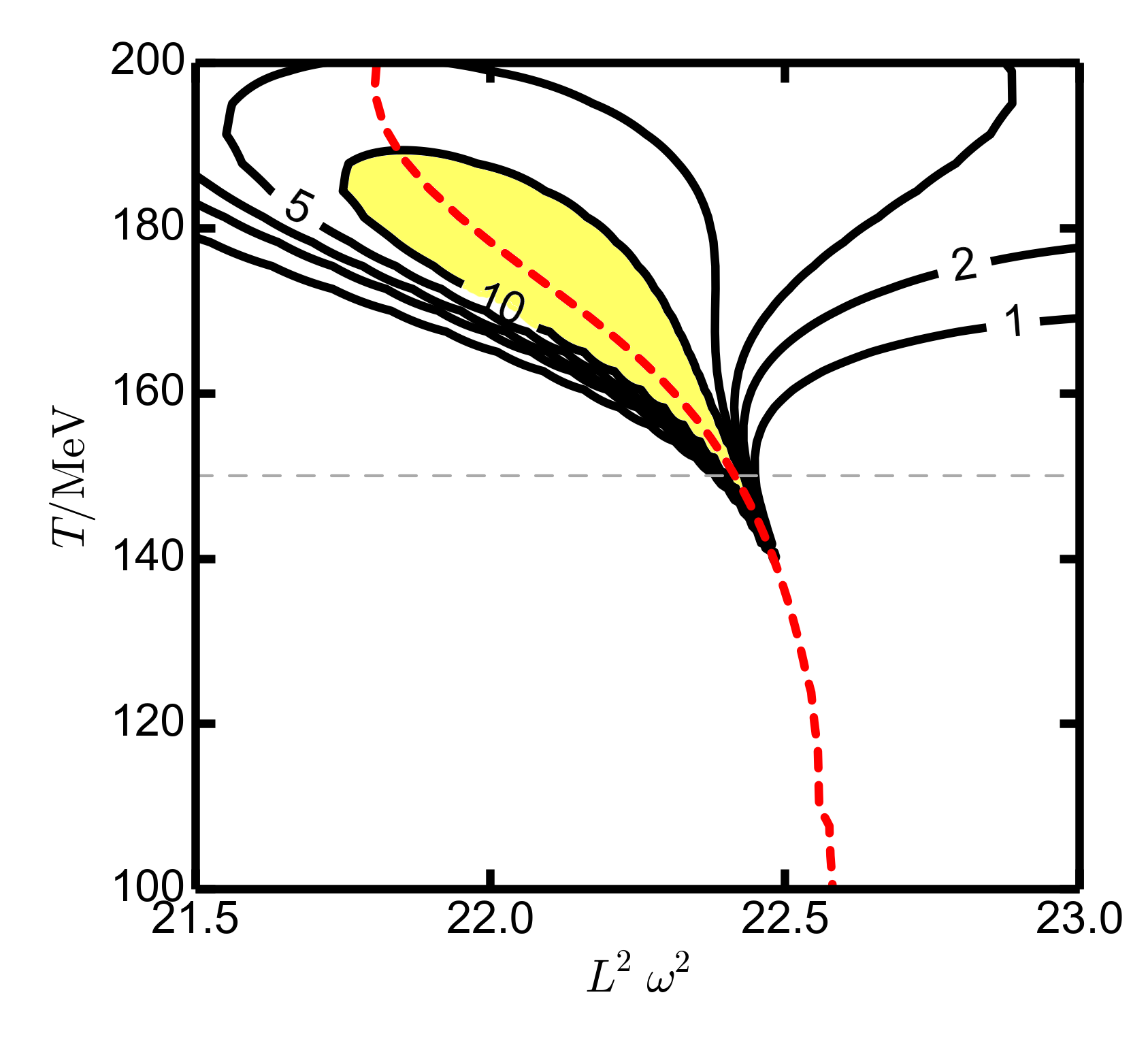}
\put(-699,90){{\huge $\Upsilon: m_{0, 1}$}}
\put(-300,90){{\huge $\Upsilon: m_0, m_1 + 5$\%}}
}
\resizebox{0.99\columnwidth}{!}{%
\includegraphics{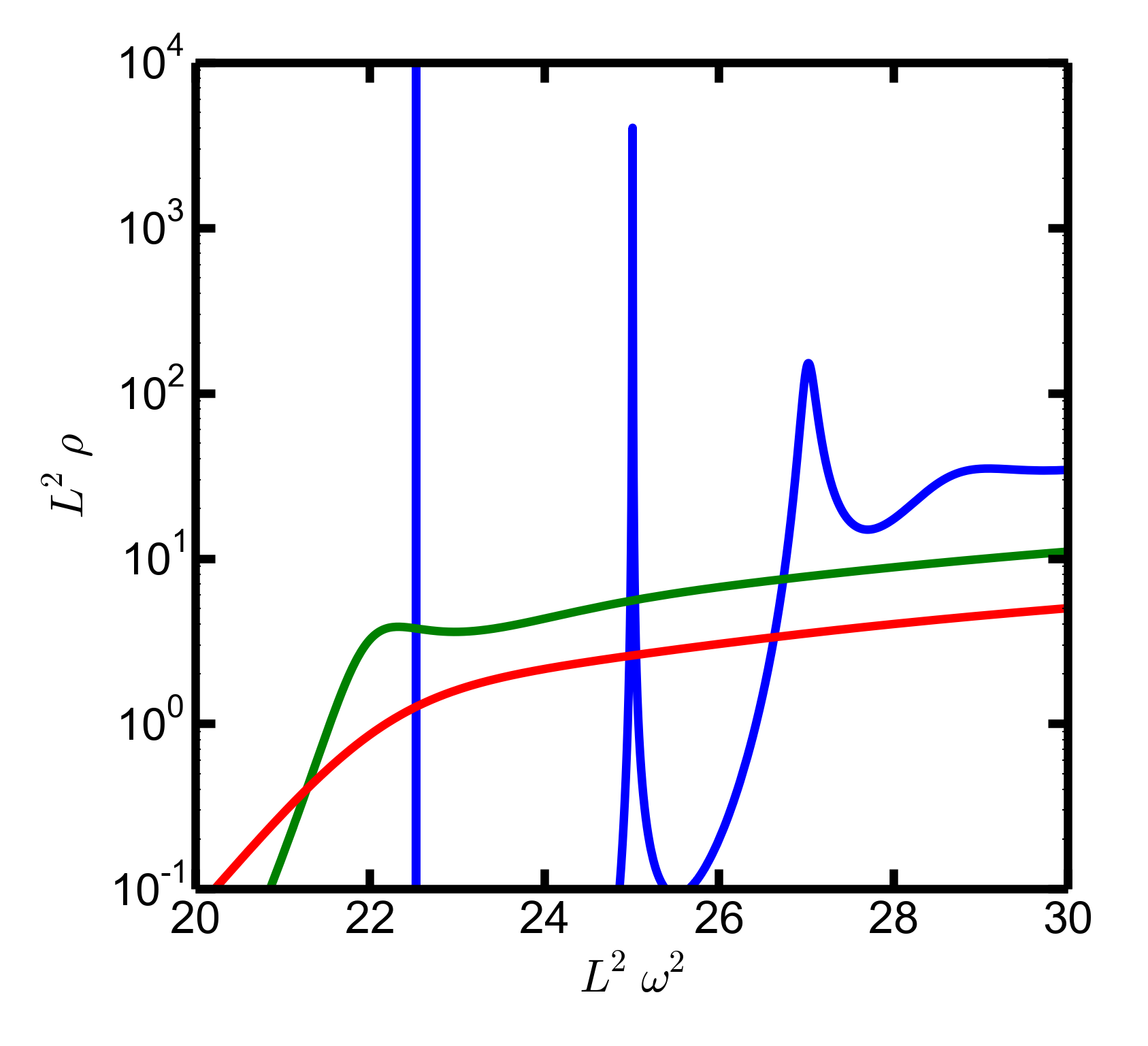}
\includegraphics{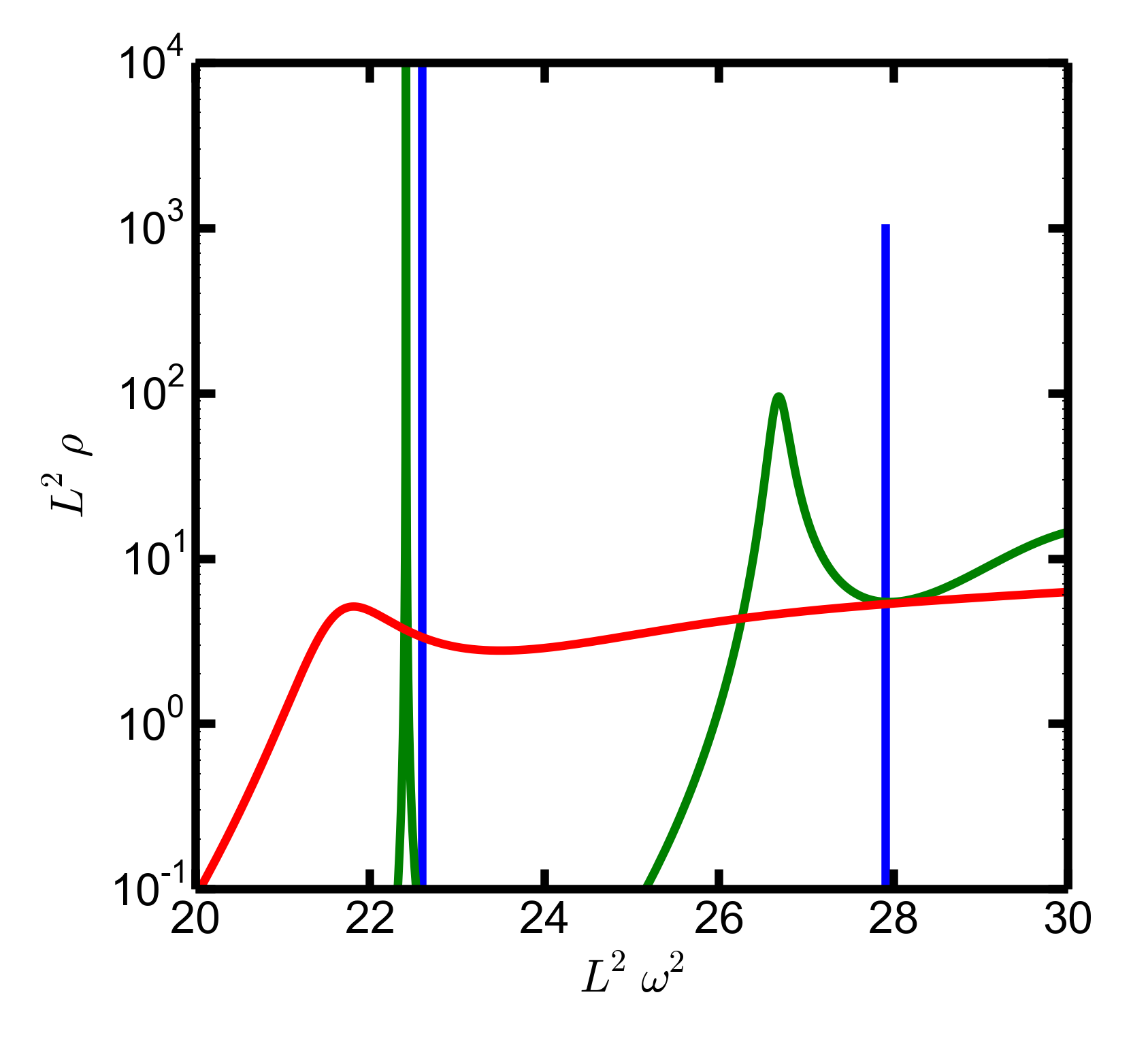}
\put(-550,299){{\huge $\Upsilon: m_{0, 1}$}}
\put(-244,299){{\huge $\Upsilon: m_0, m_1 + 5$\%}}
}
\caption{Bottomonium formation.
Top row: contour plots of the spectral functions $L^2 \rho (\omega, T)$ (the red dashed curves depict the peak position of the respective spectral function;
they terminate at $T_{melt}$; dashed horizontal lines indicate $T = 150$~MeV and as in all subsequent contour plots, the contour curve $L^2 \rho =10$ encircles the yellow area),
bottom row: spectral functions $L^2 \rho(\omega, T)$ at several temperatures
($T = 100$ (blue), 150 (green), and 200~MeV (red),
left column: for potential parameters $(a, b) = (0.6924, 4.9571)$, 
i.e.\ at such values where the $m_{0,1} (a, b)$  trajectories cross,
right column: $(a, b) = (1.3266, 4.3229)$,
i.e.\ at crossing points of the $m_0$ trajectory with the upper limit 
of the 10\% corridor of the respective 1st excitation
($T_{melt}^{{\rm g.s.}} = 234$~MeV).
\label{fig:A2}
}
\end{figure}

\begin{figure}[t]
\center
\resizebox{0.99\columnwidth}{!}{%
\includegraphics{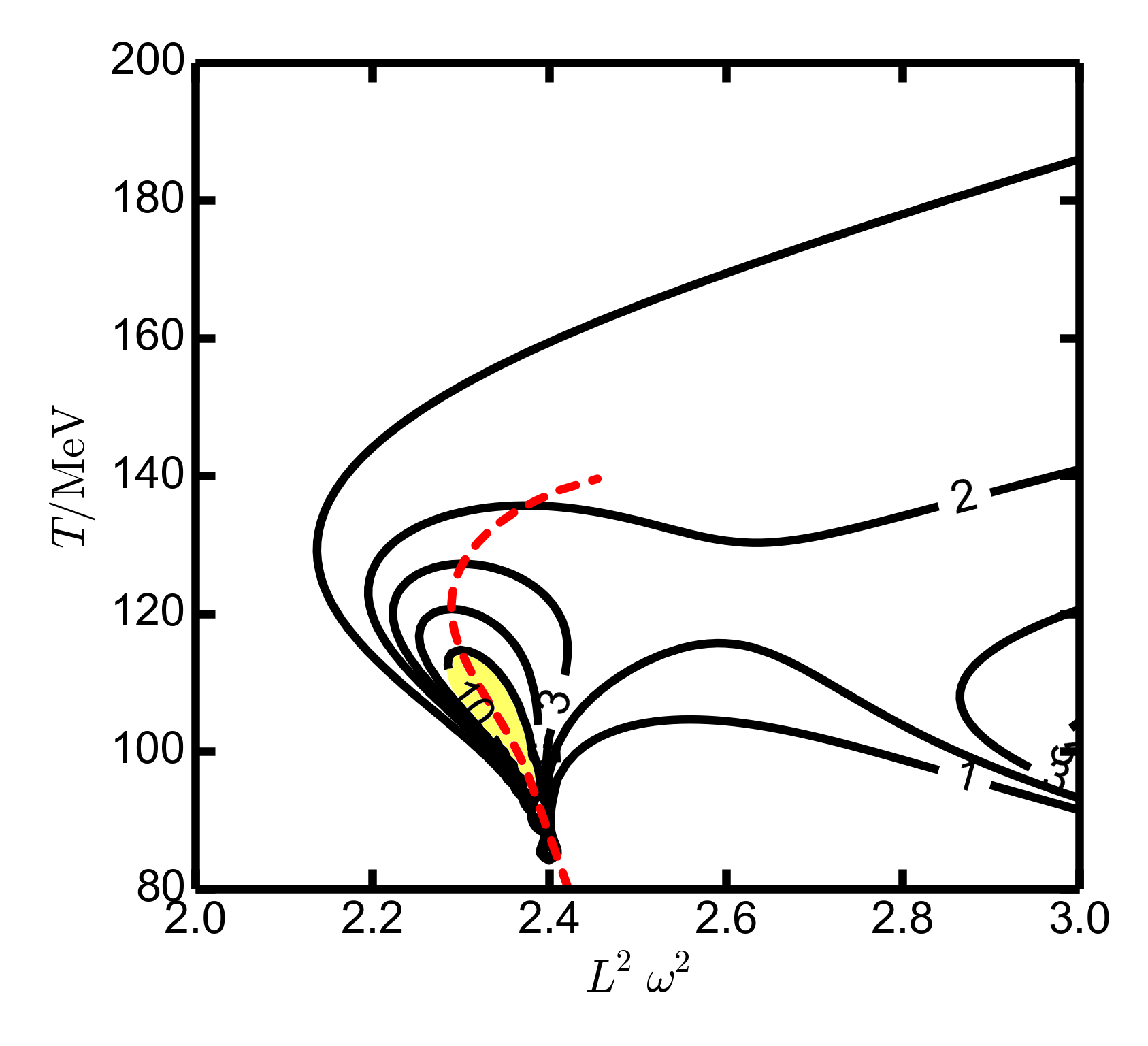}
\includegraphics{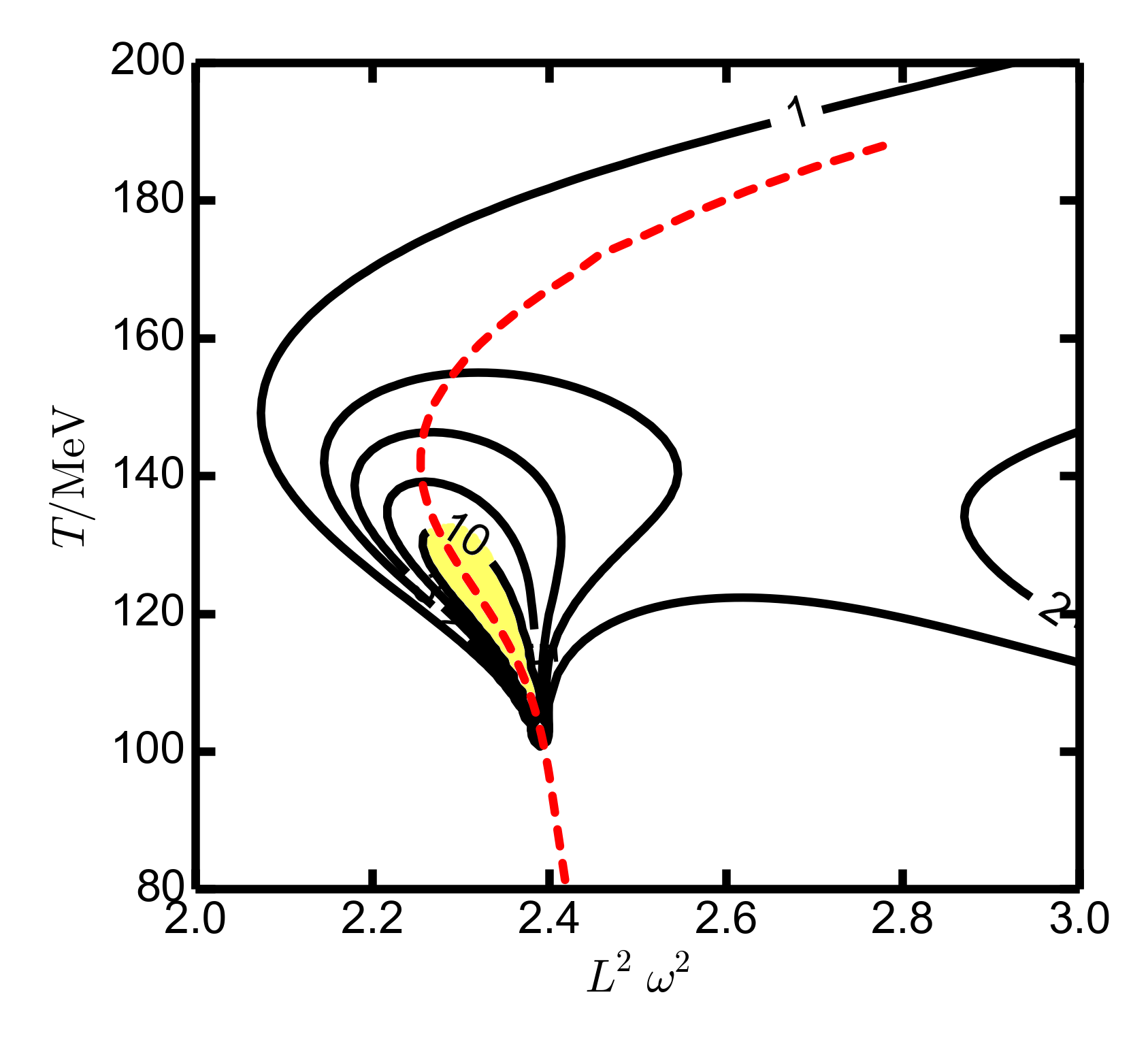}
\put(-699,275){{\huge $J/\psi: m_{0,1}$}}
\put(-189,265){{\huge $J/\psi: m_0, m_1 + 5$\%}}
}
\resizebox{0.99\columnwidth}{!}{%
\includegraphics{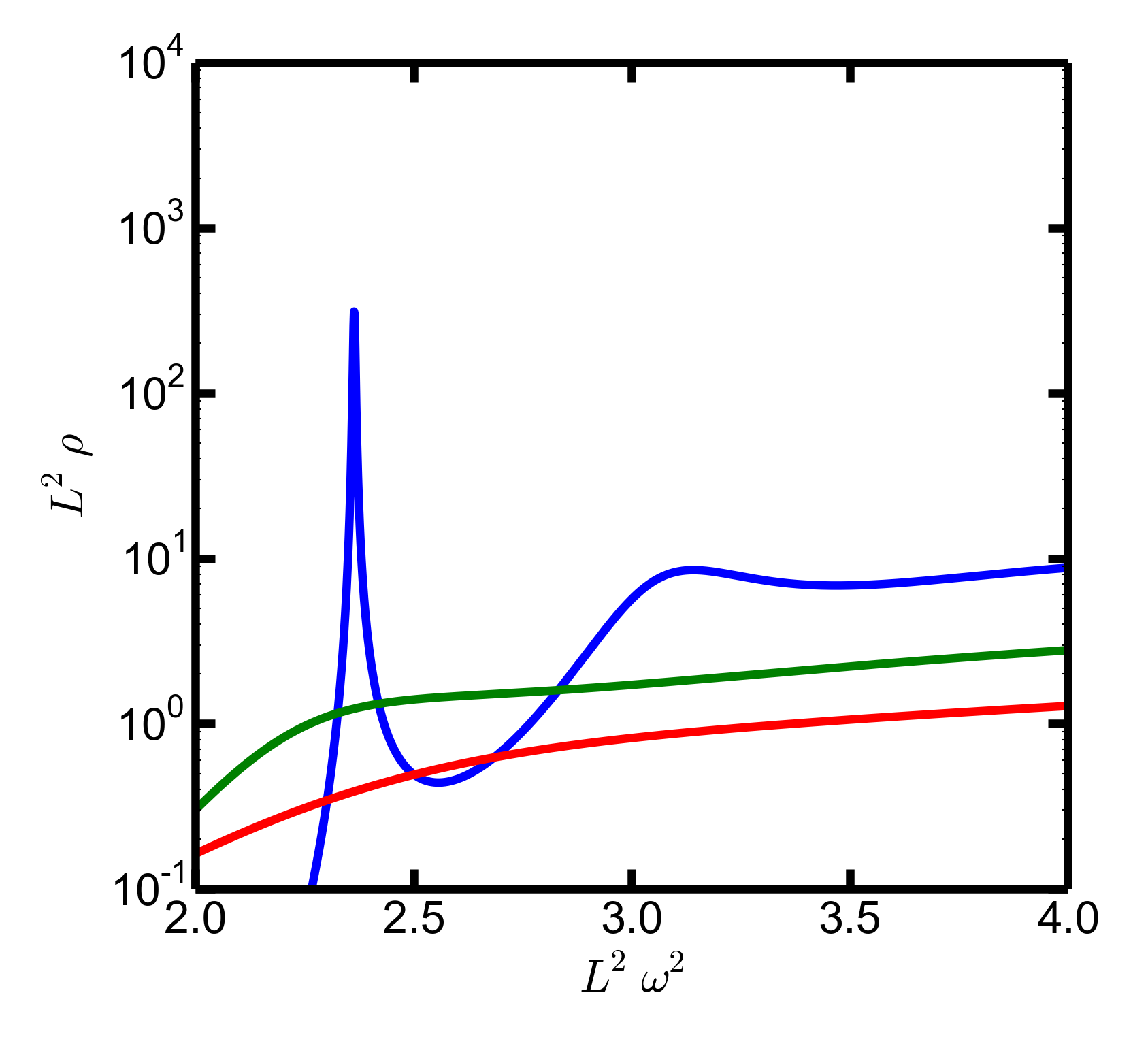}
\includegraphics{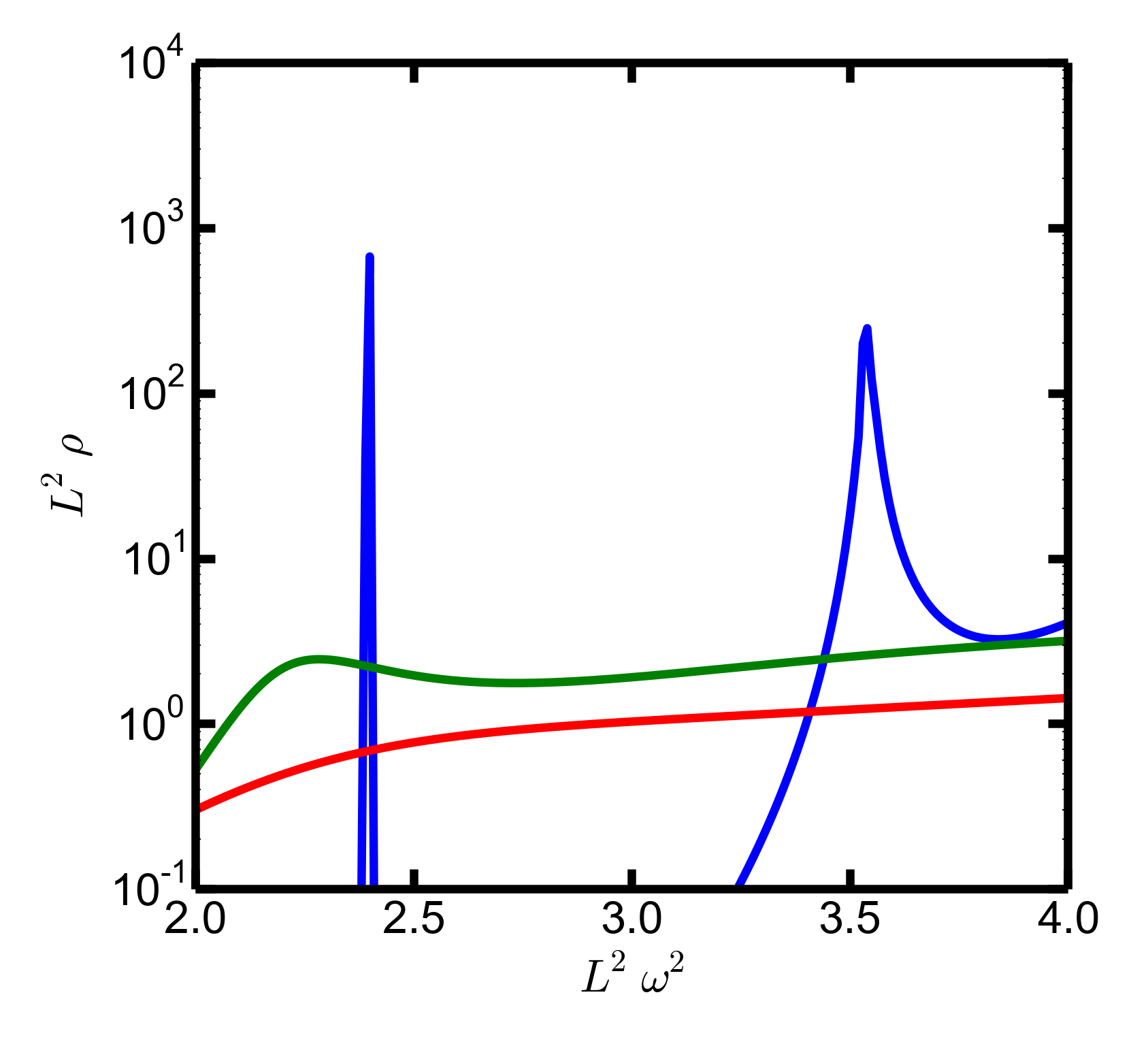}
\put(-655,299){{\huge $J/\psi: m_{0, 1}$}}
\put(-255,299){{\huge $J/\psi: m_0, m_1 + 5$\%}}
}
\caption{Charmonium.
Top row: contour plots of the spectral functions $L^2 \rho(\omega, T)$,
bottom row: spectral functions $L^2 \rho (\omega, T)$
in the energy region of g.s.\ and first excitation at several temperatures
($T = 100$ (blue), 150 (green), and 200~MeV (red)),
left column: $(a, b) = (0.2522, 0.3533)$ 
i.e.\ at such values where the $m_{0,1} (a, b)$  trajectories cross.
right column: $(a, b) = (0.338, 0.2675)$,
i.e.\ at crossing points of the $m_0$ trajectory with the upper limit 
of the 10\% corridor of the respective 1st excitation. 
\label{fig:A3}
}
 \end{figure}

In figure~\ref{fig:A1}, it looks like an accidental coincidence
that, at the crossing points of the g.s.\ and first excitation trajectories
of $J/\psi$ and $\Upsilon$, the melting temperature is 150~MeV.
In other words, in a cooling system the formation of the quarkonium 
ground state seems to start when passing the temperature of 150~MeV.
This is consistent with the claim in \cite{Andronic:2017pug} which advocates
the formation of hadron states at $T \approx 155~\mbox{MeV} \approx T_c$.
Consistency does not necessarily mean perfect agreement: The criteria for
``melting" or  ``onset of formation" are not very sharp. For instance,
\cite{Hohler:2013vca} uses as threshold value 
the relative high of the spectral function's peak 
over the smooth background for defining  ``melting". The transition to
a quasi-particle with sharp spectral function does not happen instantaneously
but within some temperature span, see top panels in
figures~\ref{fig:A2} and \ref{fig:A3}. Considering the
dynamics of the cooling system as a sequence of equilibrium states, the
spectral-function contour-plots in figures~\ref{fig:A2} and \ref{fig:A3} are suggestive: upon cooling
the strength of a hadron state is consecutively concentrated to a narrow
energy range, eventually forming the quasi-particle. 
Displaying a spectral function at a few selected temperatures, 
as in figure~\ref{fig:A4}
and bottom panels of figures~\ref{fig:A2} and \ref{fig:A3} as well, illustrates
such a feature only insufficiently but is useful for a more quantitative
account. 

Inspection of the top panels of figures~\ref{fig:A2} and \ref{fig:A3}
unravels that the temperature difference 
from $T_{melt}^{{\rm g.s.}}$ until the formation of a sharp quasi-particle
state is quite large. Sharp quasi-particles can be identified by the squeezed
contour lines which eventually coincide nearly with the peak position of the
spectral functions depicted
by the red dashed curves in top panels of figures~\ref{fig:A2} and \ref{fig:A3}. 
Keeping the quarkonia ground state masses $m_0$ and allowing artificially for a somewhat larger value
of the first excited state $m_1$ moves the quarkonia formation temperatures
to larger values, in particular for $\Upsilon$, see right panels in
figure~\ref{fig:A2}. In such a way, the quasi-particle formation temperature
$T_{form}^{\Upsilon (1S)} \approx T_c$
copes with the claim in
\cite{Andronic:2017pug} of hadron formation at $T_c$. The $J/\psi$,
in contrast, would be formed at $T_{form}^{J/\psi} < T_c$
(see figure~\ref{fig:A3}) in conflict with the advertisement 
of \cite{Andronic:2017pug}. Section \ref{sect:three_param} 
provides a potential ansatz $U_0 (z; \vec p \,)$ which accomplishes $T_{form}^{J/\psi} \approx T_c$.

To understand why the $J/\psi$ ($\Upsilon$) reacts so sluggishly (violently)
on a modification of $m_1$ while keeping $m_0$ fixed, we mention that
the parameter $\hat a$ in (\ref{eq:10}) changes by 33\% (92\%, i.e.\ a factor of nearly two)
upon a 5\% increase of $m_1$,\footnote{Due to the non-linearity of the
$J^{PC} = 1^{--}$ bottomonium Regge trajectory, the energy of
$m_1 + 5$\% is between the $3^3 S_1/\Upsilon (3S)$ and
$4^3 S_1/\Upsilon (4S)$ states. For charmonium, in contrast,
$m_1 + 5$\% is well below the $3^3 S_1/\psi (4040)$ state, cf.\
\cite{Ebert:2011jc}.}
which is to been seen in connection
with the curvature $8 \hat a^2$ of $U_0$ 
at the minimum $z_{min} = \left( 3/(4 \hat a^2) \right)^{1/4}$. The more the
potential $U_0$ is squeezed by parameter variation, 
e.g.\ by larger values of $\hat a$, 
the less is the temperature sensitivity of $U_T$, 
see figure~2 in \cite{Zollner:2020cxb} and figure~\ref{fig:Braga3} below.
At the origin of these differences is the ratio of $m_1^2/m_0^2$ which is
1.42 for $J/\psi$ and 1.12 for $\Upsilon$, respectively. It enters the
scaled potential (\ref{eq:10})
$U_0(\hat z \equiv m_0 z, \zeta \equiv m_1^2/m_0^2 - 1) / m_0^2 
= \frac34 \hat z^{-2} + \frac{1}{16} \zeta^2 \hat z^2 - \zeta + 1$
as solely parameter.

A second issue refers to the formation of excited states. It seems to be a generic feature of the holographic model class
considered here that higher excited states would form at lower temperatures than the respective g.s., in particular
$T_{form}^{{\rm g.s.}} > T_{form}^{{\rm 1st}} > T_{form}^{{\rm 2nd}} \cdots$,
see bottom panels in figures~\ref{fig:A2} and \ref{fig:A3}.
The conjecture of \cite{Andronic:2017pug}, in contrast, advocates 
$T_{form}^{{\rm g.s.}} \approx T_{form}^{{\rm 1st}}$. This feature is to be seen
in relation to the considered ansatz of $U_0 (z; a, b)$ with the IR behavior
$\propto z^2$: a much steeper increase of $U_0$ at larger values of $z$
would concentrate the melting temperatures in a narrow corridor.

\begin{figure}[t]
\center
\resizebox{0.99\columnwidth}{!}{%
\includegraphics{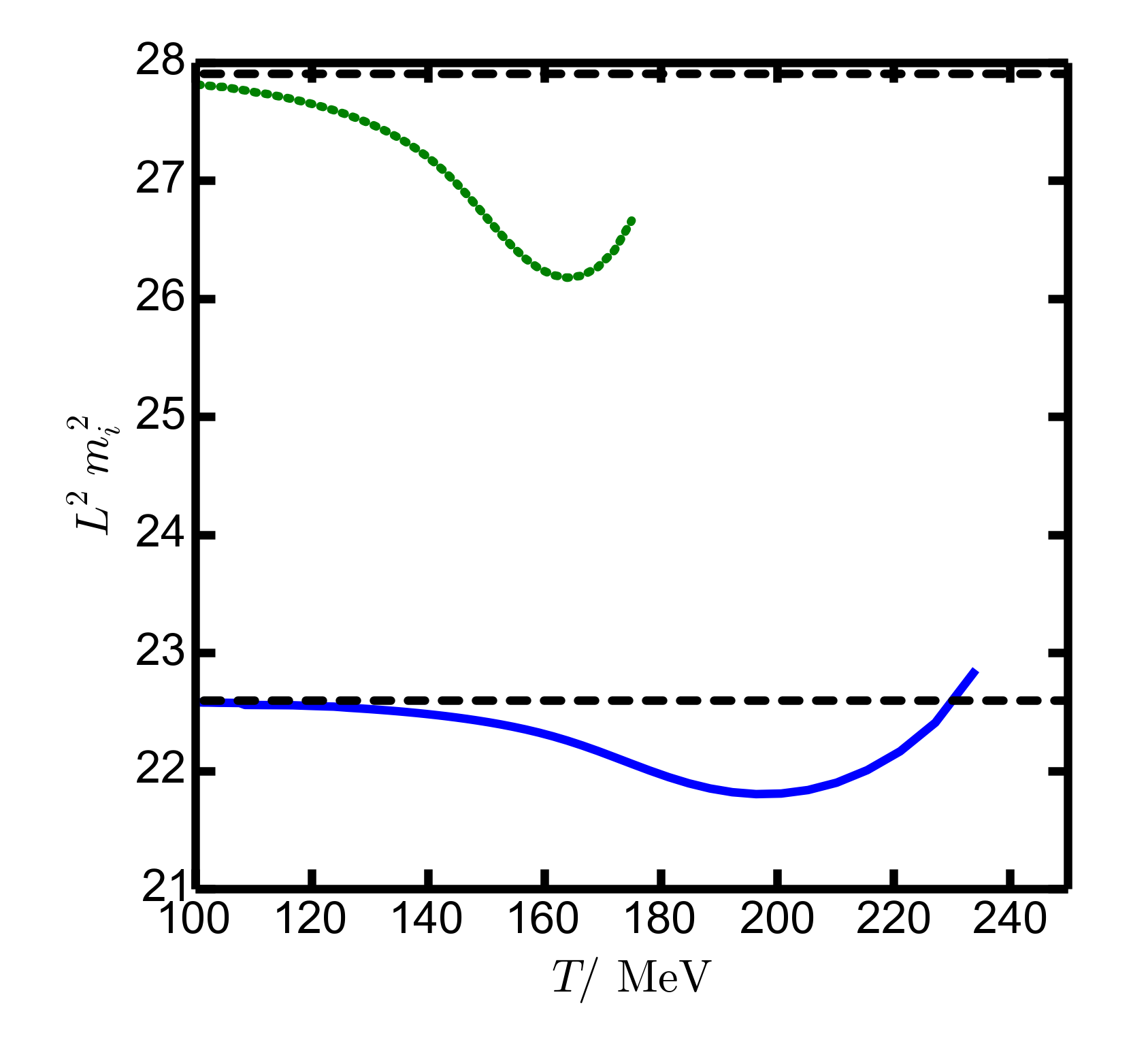}
\includegraphics{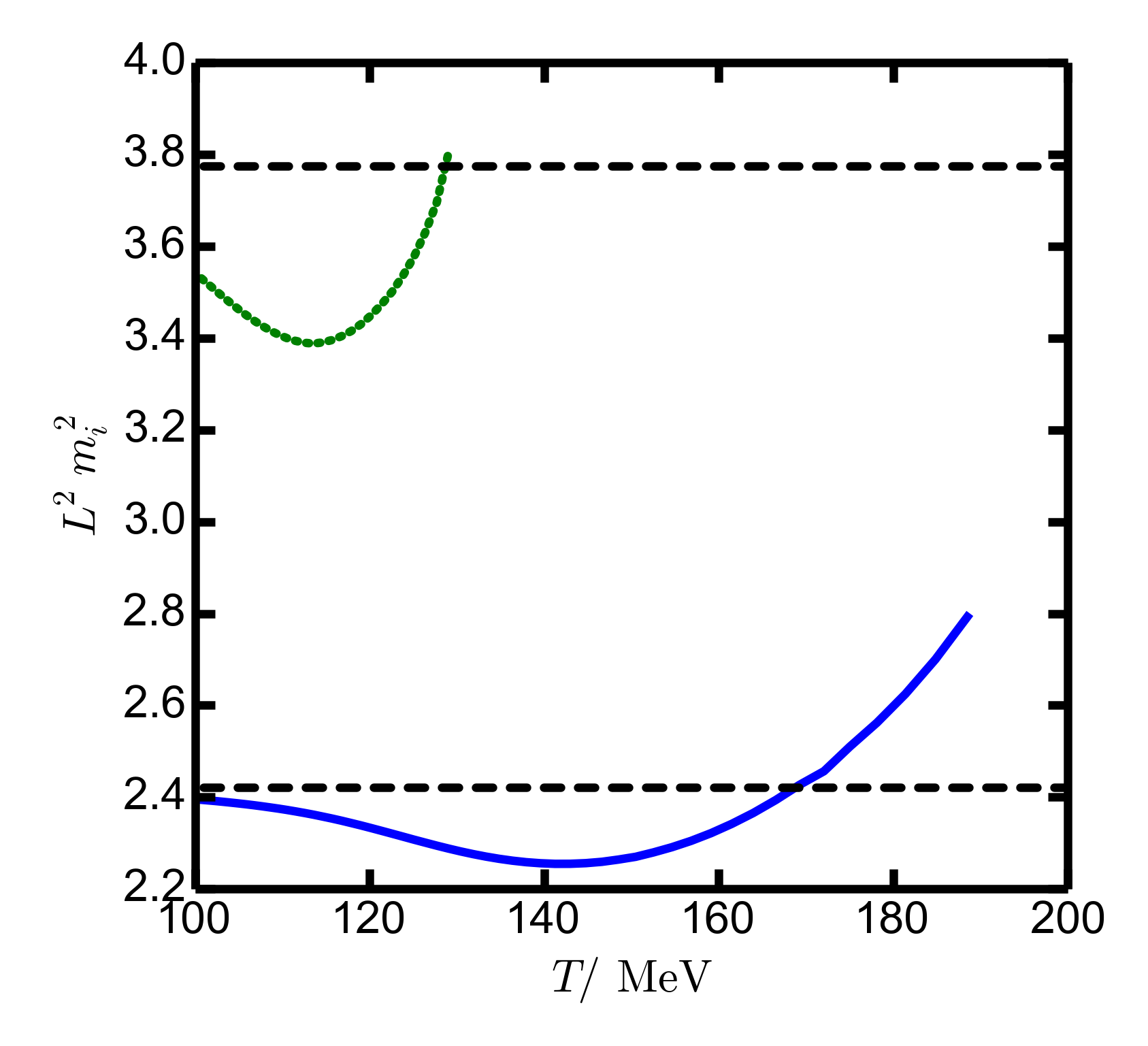}
\put(-700,205){{\huge $\Upsilon$}}
\put(-300,205){{\huge $J/\psi$}}
}
\caption{Positions of the peaks of the spectral functions 
of $\Upsilon$ (left panel) and  $J/\psi$ (right panel) as a function of 
temperature. The right end points of the solid curves
define $T_{melt}$ fore the g.s.\ (lower blue solid curves) and
1st excitation (upper green dotted curves). The dashed lines
depict the masses squared at $T = 0$. The difference of the solid or dotted curves to
the dashed lines is termed ``thermal mass shift" squared.
For $(a, b) = (1.3266, 4.3229)\vert_{\Upsilon}$
and $(0.338, 0.2675)\vert_{J/\psi}$,
as in the right columns of figures~\ref{fig:A2} and \ref{fig:A3}. 
\label{fig:A5}
}
 \end{figure}

Besides the ansatz (\ref{eq:10}) facilitates a sequential quarkonium formation upon decreasing temperature,
$T_{form}^{{\rm g.s.}} > T_{form}^{{\rm 1st}} > T_{form}^{{\rm 2nd}}$ etc.,
it allows potentially for a some flavor dependence, e.g.\ 
$T_{form}^{\Upsilon(1S)} > T_{form}^{J/\psi}$.
The thermal mass shifts have a non-trivial temperature dependence
as evidenced in figure~\ref{fig:A5}. Such thermal mass shifts 
are employed in \cite{Brambilla:2019tpt} to pin down 
the heavy-quark (HQ) transport coefficient
$\gamma$ which can be considered as the dispersive counterpart
of the HQ momentum diffusion coefficient 
$\kappa = 2 T^3 /(D T)$, where $D$ stands for the HQ spatial diffusion
coefficient. Reference \cite{Rothkopf:2019ipj} stresses 
a seemingly tension within
previous holographic results \cite{Braga:2017bml}, where {\sl positive} mass shifts
are reported, in contrast to {\sl negative} shifts, e.g.\ in  \cite{Fujita:2009ca}.
Our set-up resolves qualitatively that issue since, depending on the considered
temperature, the thermal mass shift can be negative or positive,
see figure~\ref{fig:A5}.
One should note, however, that our thermal mass shifts of $J/\psi$ and $\Upsilon$ 
are larger than the lattice QCD-based values 
quoted in \cite{Brambilla:2019tpt,Larsen:2019zqv}.
 
Finally, let us remind that the two-parameter ansatz (\ref{eq:10})
is appealing since it allows for analytic solutions w.r.t.\ the excitation
spectrum and an easy overview on the parameter dependencies.
However, already the authors of \cite{Grigoryan:2010pj,Hohler:2013vca}
promoted (\ref{eq:10}) to a ``shift and dip potential" to catch more
properties of the $J/\psi$ states than only masses.

\section{Three-parameter potential with dip -- charmonium formation}
\label{sect:three_param}

The two-parameter potential $U_0 (z; a, b)$  from (\ref{eq:10}) with realistic
values of $a(m_{0, 1})$ and $b(m_{0, 1)}$ facilitates $J/\psi$ formation at too low
temperatures. This failure can be repaired by turning to more appropriate
parameterizations. For instance, \cite{Grigoryan:2010pj,Hohler:2013vca}
proposed a four-parameter ``dip and shift potential" which allows for
$J/\psi$ melting temperatures significantly above $T_c$, as also the
construction in \cite{Braga:2015lck,Braga:2016wkm,Fujita:2009wc,Fujita:2009ca,Braga:2017bml} deploying three
parameters.
The essence is a dip in $U_0 (z; \vec p \,)$ which holds together
the spectral strength despite large temperatures. 
Here, we consider such an option. The difference
to previous work is the use of the dynamical background related to QCD,
as described in Appendix \ref{App:B}.

The construction of a particular three-parameter potential
$U_0 (z; M, k, \Gamma)$ is as follows.
Use $A_0 (z) = - 2 \log z/L$ and 
$\phi_m (z; M, k, \Gamma) = k^2 z^2 + M z + \tanh x $ \cite{Braga:2018zlu}
with $x \equiv 1/ M z - k/\sqrt{\Gamma}$.
Due to
$U_0 = \frac12 (\frac12 A_0' - \phi_m')' + \frac14 (\frac12 A_0' - \phi_m')^2$
(see (\ref{eq:8}, \ref{eq:9})) the potential is given by
\begin{eqnarray} \label{eq:Braga_U0}
U_0 (z; M, k, \Gamma) &=& 
\frac{3}{2 z^2} + \frac14 M^2+ k^4 z^2 
+ \frac{M}{2 z} + k^2 M z \nonumber \\
&-& \left( \frac{3}{2M z^3} + \frac{1}{2 z^2} + \frac{k^2}{M z}
\right) \frac{1}{\cosh^2 x} \nonumber \\
&+& \frac{1}{4 M^2 z^4} \left[ 4 \sinh x \, \cosh x + 1 \right]  
\frac{1}{\cosh^4 x} .
\end{eqnarray}
The first three terms in the top line suggest a correspondence
$M \widehat{=} 4 \sqrt{b} / L$ and $k \widehat{=} \sqrt{a} / L$ by a comparison
with (\ref{eq:10}), while the next two terms cause some modification
of (\ref{eq:10}) at intermediate values of $z$.
The second line of (\ref{eq:Braga_U0}) is essentially responsible for
the dip -- somewhat modified by terms in the third line.
The dip position is determined to a large extent by the
$1/\cosh^2$ term which peaks at $z = \sqrt{\Gamma} /(k M)$;
the $\sinh$ term in the third line shifts the dip tip to smaller values
of $z$. The UV and IR asymptotics are the same as for the potential (\ref{eq:10}).
The dip position and the dip depth are interrelated, in contrast to the construction
in  \cite{Grigoryan:2010pj,Hohler:2013vca}.
 
The potential (\ref{eq:Braga_U0}) 
might exhibit some non-trivial local structures as a function of $z$ for
particular parameters. Reference \cite{Braga:2018zlu}
advocates the optimum parameters 
$M = 2.2$ GeV (representing a mass scale of non-hadronic decays), 
$k = 1.2$ GeV (representing the quark mass) and 
$\sqrt{\Gamma} = 0.55$ GeV (representing the $Q \bar Q$ string tension) 
to yield the $J/\psi$ ($\psi'$)
mass of 2.943 (3.959)~GeV) and the decay width of 399 (255)~MeV.
Note the resulting overestimated level spacing quantified by
$m_1^2 / m_0^2 = 1.81$, in contrast to the PDG value of 1.42,
when deploying these parameters.

\begin{figure}[tb]
\center
\resizebox{0.99\columnwidth}{!}{%
\includegraphics{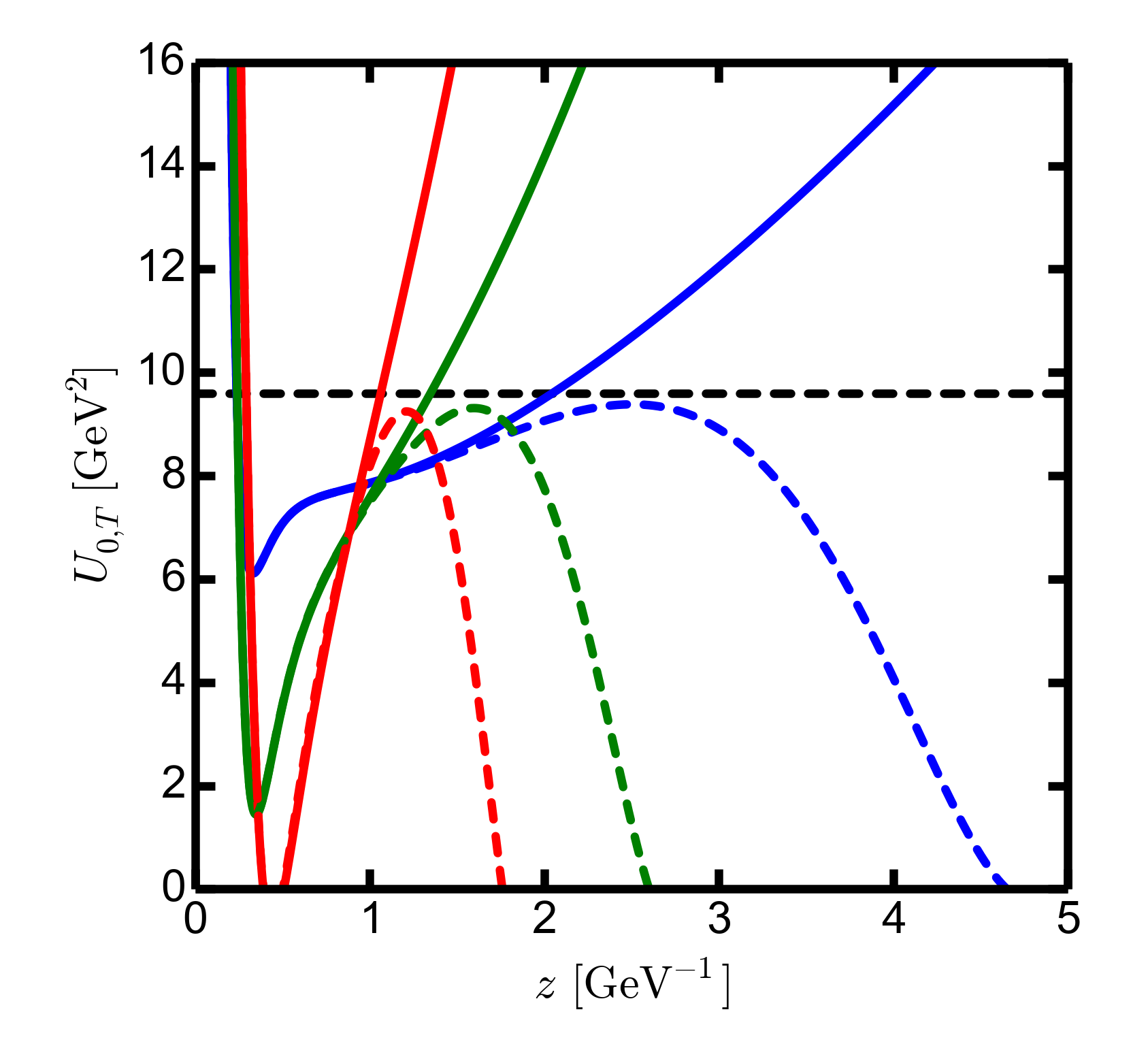}
\includegraphics{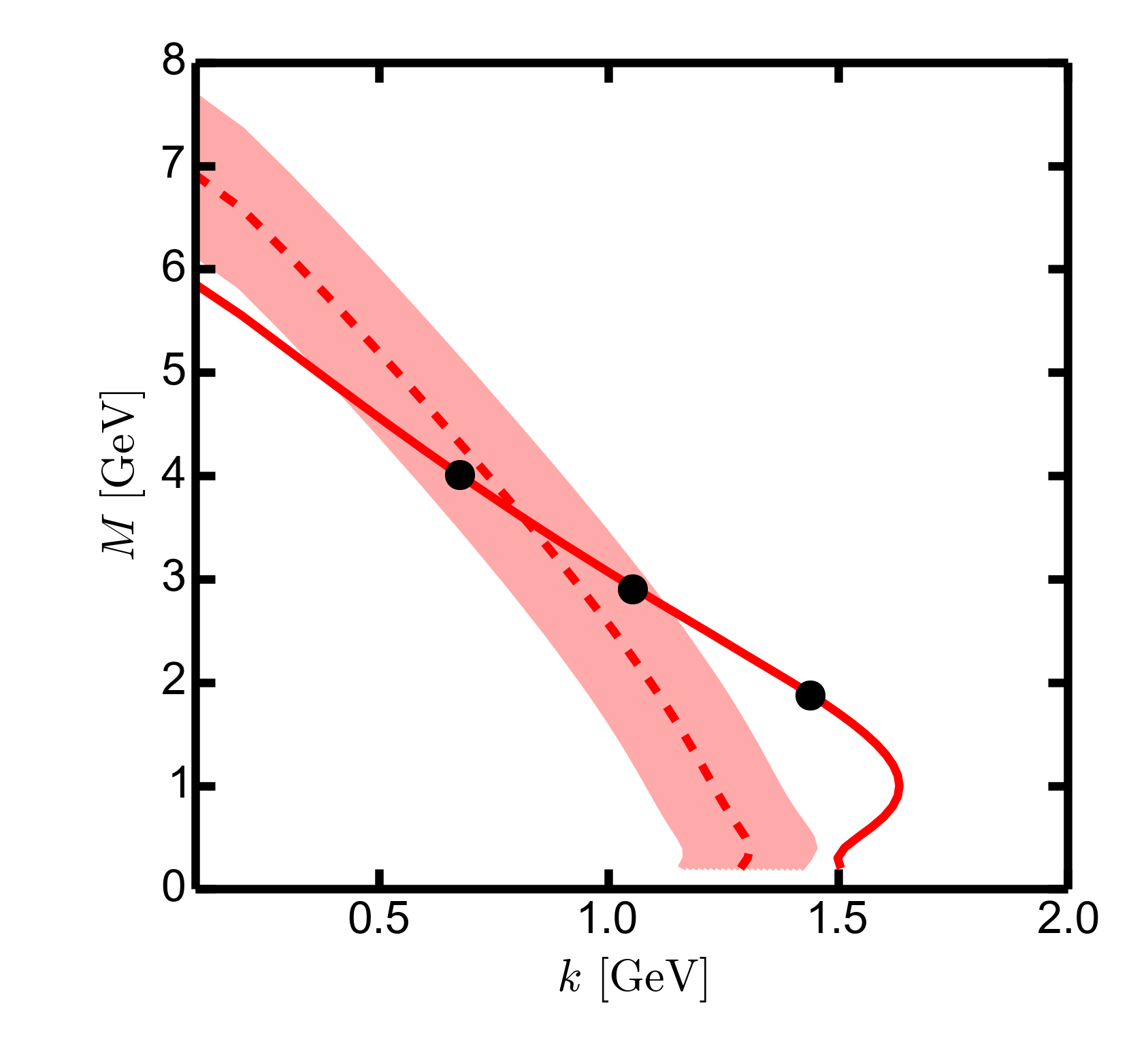}
\put(-480,300){{\Huge $U_0$}}
\put(-480,188){{\Huge $U_T$}}
\put(-545,150){{\huge {\color{blue} $T =100$ MeV}}}
\put(-590,125){{\huge {\color{green}$150$}}}
\put(-635,100){{\huge {\color{red} $200$}}}
\put(-200,222){{\huge {\color{red}$J/\psi$}}}
}
\caption{Left panel: The potential $U_0$ from 
(\ref{eq:Braga_U0}) (upper fat solid curves)
and the resulting $U_T$ from 
(\ref{eq:7}, \ref{eq:G}, \ref{eq:8}, \ref{eq:9}) (lower dashed curves)
as a function of $z$ for $\sqrt{\Gamma} = 1.5$~GeV and 
values $(k, M) = (0.676, 4.007)$ (blue), $(1.053, 2.90)$ (green) and
$(1.44, 1.87)$ (red). 
The dashed horizontal line depicts the
$J/ \psi$ g.s.\ mass squared $m_0^2$
from $U_0$ which is the same for all three parameter selections.  
Right panel: The $m_{0, 1}$ trajectories with PDG values
in the $k$-$M$ plane at $\sqrt{\Gamma} = 1.5$~GeV. 
The $m_1 \pm 5$\% corridor is depicted as colored band.
The three parameter pairs $(k, M)$ of the left panel are shown by bullets.
\label{fig:Braga3}
}
\end{figure}
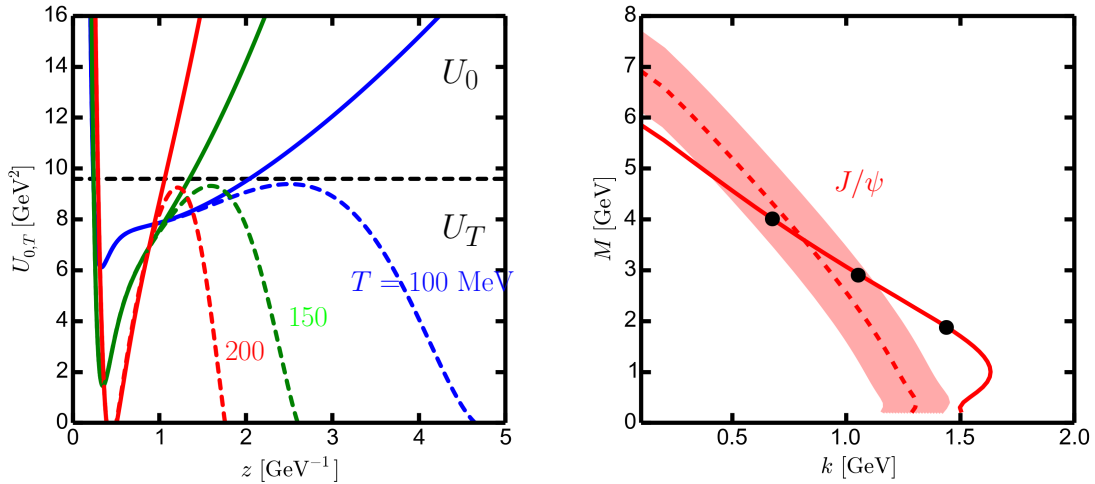

Completely analog to the two-parameter potential (\ref{eq:10}),
increasing the parameter $k$ at $M \approx const$, the
potential (\ref{eq:Braga_U0}) is squeezed and becomes deeper.
Analogously, decreasing the parameter $\sqrt{\Gamma}$
at constant values of $k$ and $M$ lets drop the absolute
minimum of $U_0$.
One may select such parameter pairs of $(k, M)$ at constant
$\sqrt{\Gamma}$ to keep the g.s.\ mass $m_0$ constant,
see the horizontal dashed line in left panel of figure~\ref{fig:Braga3}.
Due to the squeezing of the potential, the interior (left) part
is less influenced when imposing a horizon at $z_H$,
where $U_T(z = z_H) = 0$ is facilitated according to (\ref{eq:7}). 
As a result, the more the potential
is squeezed the smaller values of $z_H$ are allowed to hold the 
$J/\psi$ prior to melting. That is the very reason which forces us to enlarge
the parameter $k$ (or $a$ in (\ref{eq:10})) to achieve quarkonium
formation at sufficiently high temperatures in agreement with the
perspective put forward in \cite{Andronic:2017pug}. The dip in the
potential (\ref{eq:Braga_U0}) is useful in that respect since enlarging
the parameter $a$ in the flat potential (\ref{eq:10}) influenced the
quarkonium formation in a less effective manner for $J/\psi$.
Let us emphasize that we put more weight on the g.s.\ mass $m_0$ 
(see fat solid curve in the right panel of figure~\ref{fig:Braga3})
as the representative of the quark mass, while we relaxed the constraint
on the excited state $m_1$ to be in a realistic range
(see dashed curve and colored band in the right panel of figure~\ref{fig:Braga3}),
thus following the rationale in \cite{Grigoryan:2010pj}.

\begin{figure}[tb]
\center
\resizebox{0.99\columnwidth}{!}{%
\includegraphics{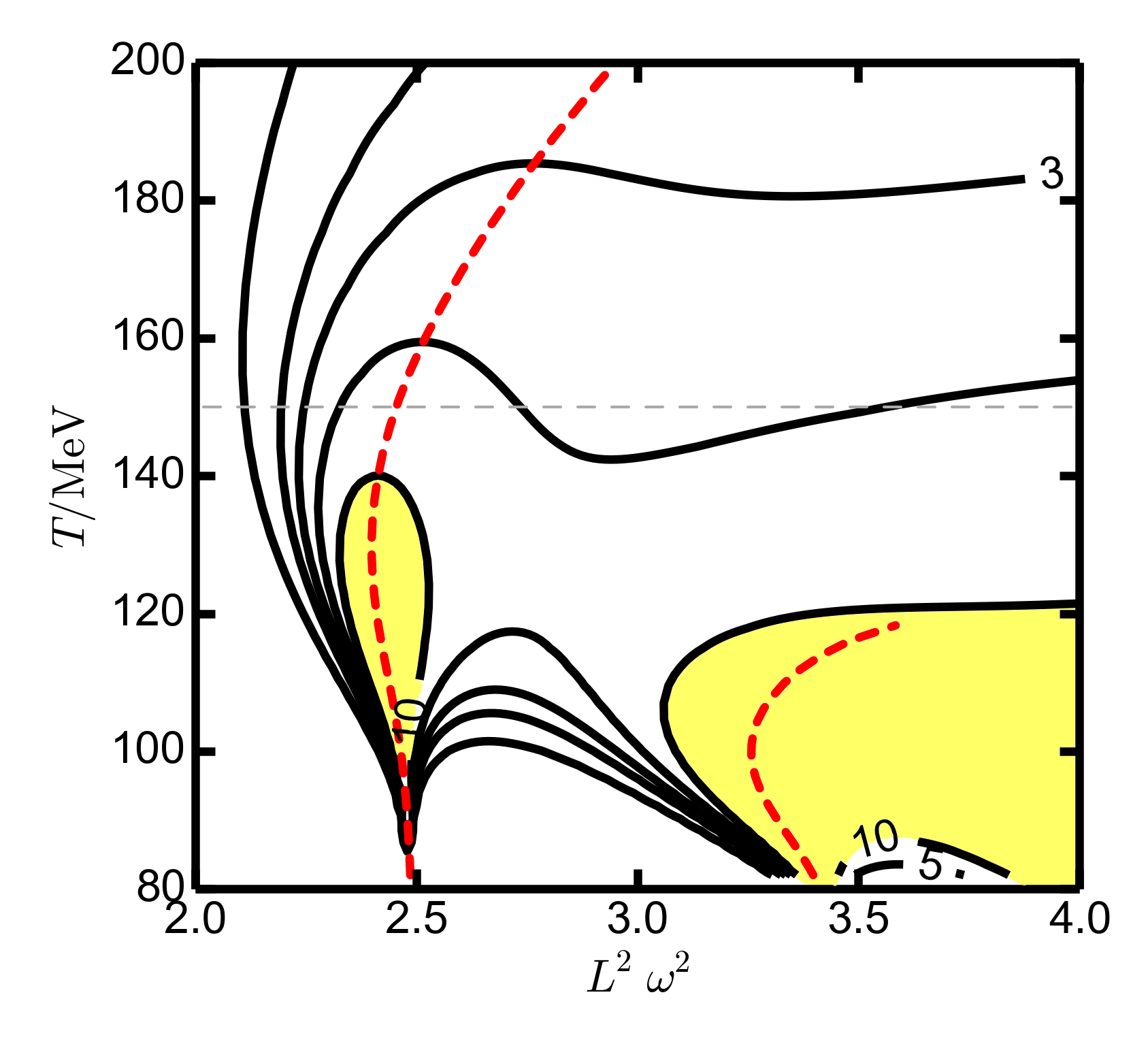}
\includegraphics{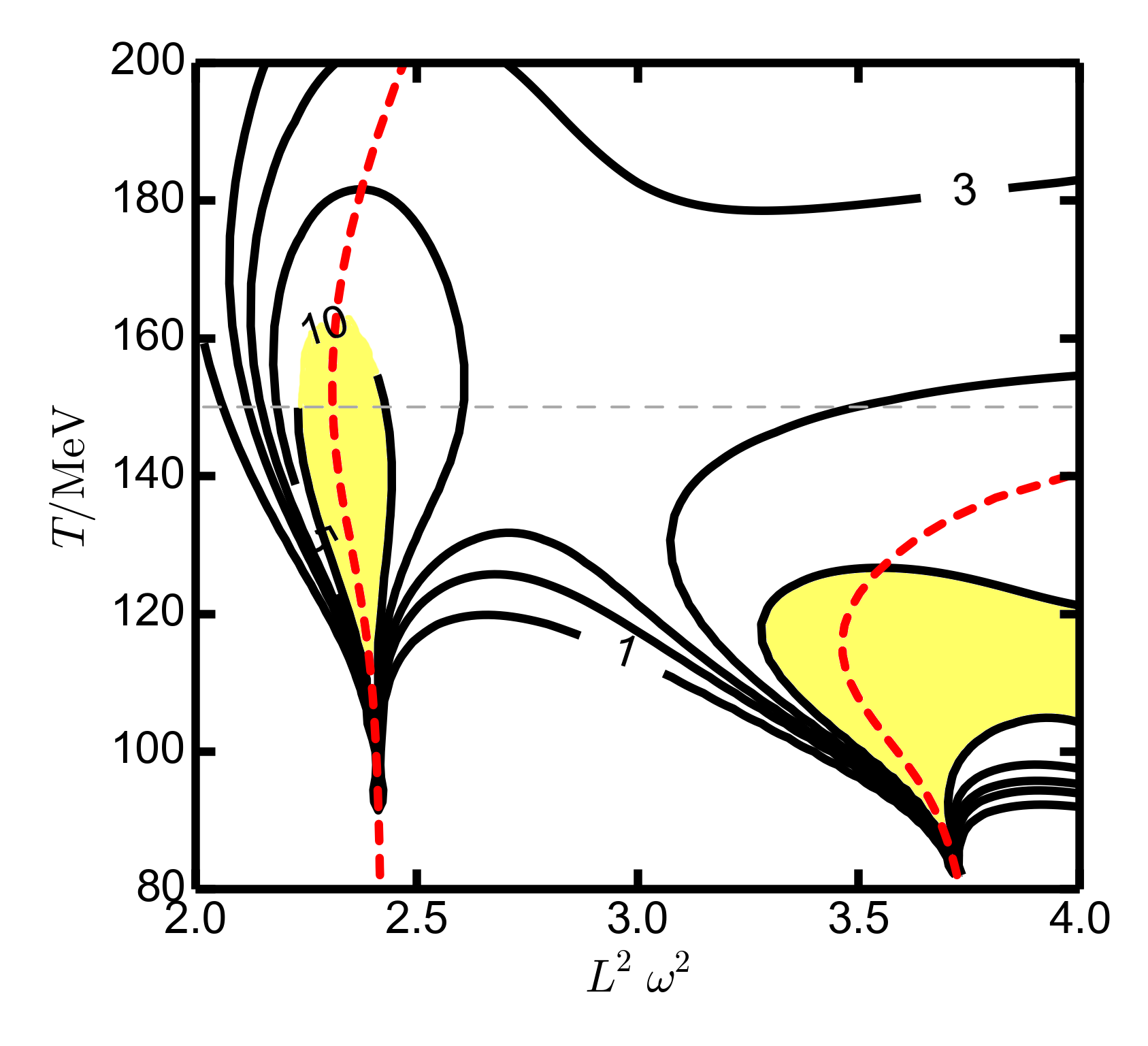}
\put(-739,221){{\Huge -----------------------------------------}}
\put(-338,221){{\Huge -----------------------------------------}}
\put(-588,265){{\huge $J/\psi: m_{0, 1}$}}
\put(-240,265){{\huge $J/\psi: m_0, m_1 + 5$\%}}
}
\resizebox{0.99\columnwidth}{!}{%
\includegraphics{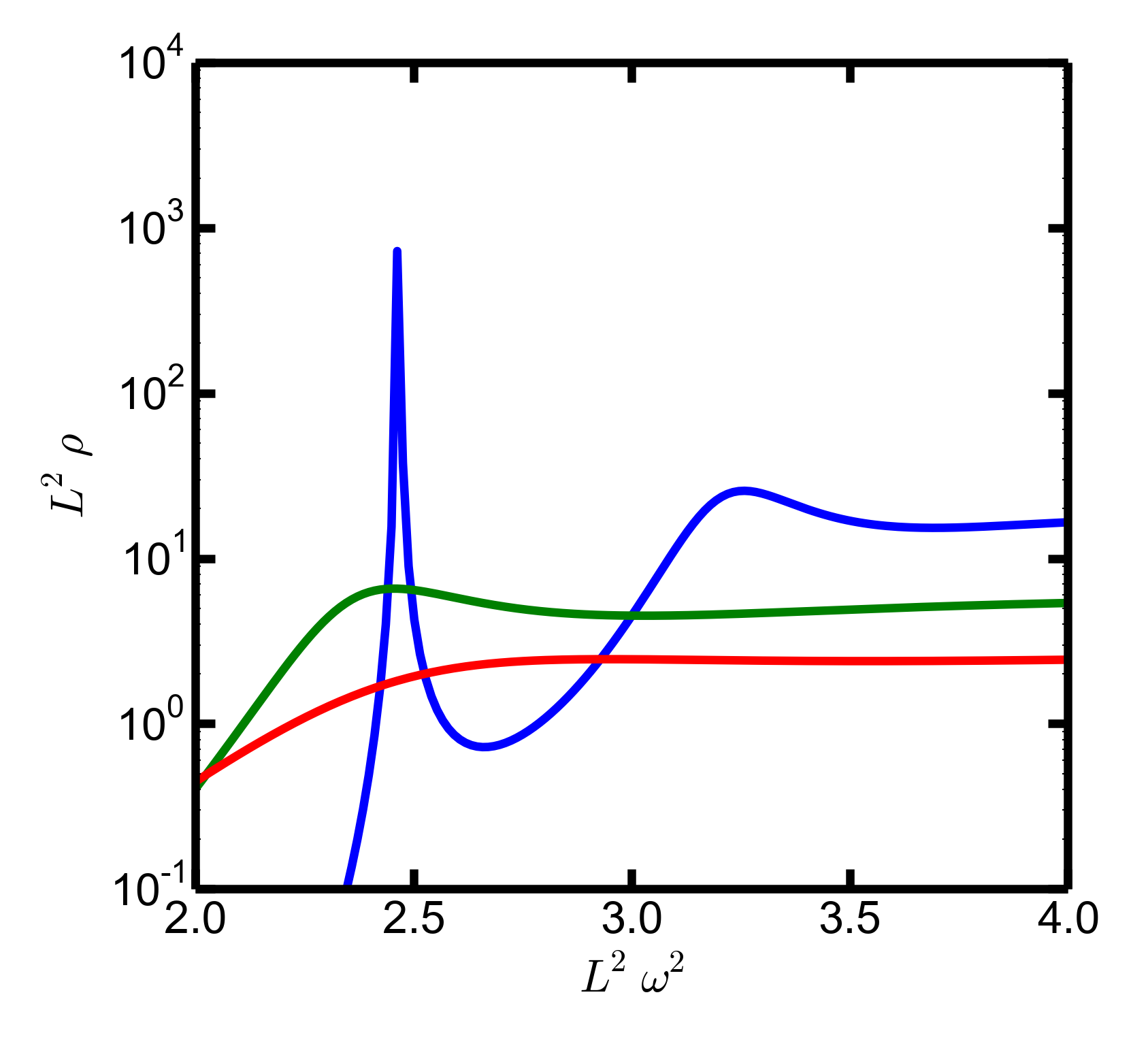}
\includegraphics{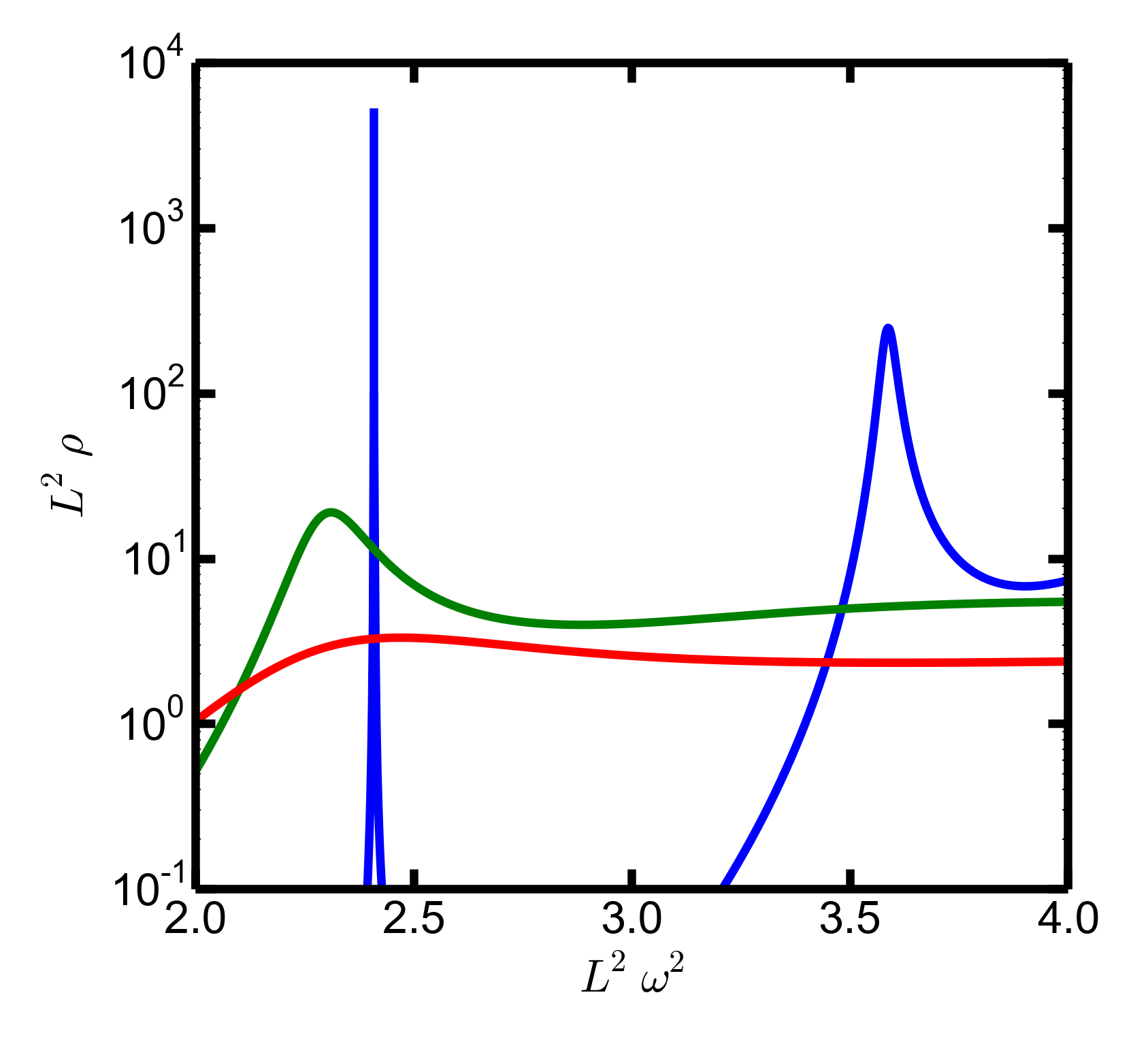}
\put(-588,300){{\huge $J/\psi: m_{0, 1}$}}
\put(-240,300){{\huge $J/\psi: m_0, m_1 + 5$\%}}
}
\caption{Charmonium formation 
in the three-parameter potential (\ref{eq:Braga_U0}).
Top row: contour plots of the spectral functions
(the red dashed red curves depict the peak positions of the 
spectral functions),
bottom row: spectral functions
for a few selected values of the temperature, 
$T = 100$ (blue), 150 (green), and 200~MeV (red),
left column: $(\sqrt{\Gamma}, k, M) = (1.0, 0.825, 3.656)$~GeV
(yielding PDG values of $m_{0, 1}$, $T_{melt}^{{\rm g.s.}} = 228$~MeV),
right column: $(\sqrt{\Gamma}, k, M) = (1.0, 0.9818, 3.1398)$~GeV
(yielding $m_0$ and $m_1 + 5$\%, $T_{melt}^{{\rm g.s.}} = 361$~MeV). 
\label{fig:Braga1}
}
\end{figure}

Having $U_0$ at our disposal we proceed as in Section \ref{sect:two_param}.
Contour plots of the $J/\psi$ spectral function are exhibited in
the top panels of figure~\ref{fig:Braga1}. 
One observes again the tendency of charmonium
formation as narrow corridor of contour lines at too low temperatures
for parameters delivering exactly the PDG values of $m_{0, 1}$,
see left top panel.
This is quantified by the spectral functions shown 
in the left bottom panel of figure~\ref{fig:Braga1}
which display only a broad peak at $T = 150$~MeV.
Modifying the parameters such to catch $m_0$ and $m_1 + 5$\%
DPG values improves significantly the approach to charmonium formation
near $T_c$, see right panels of figure~\ref{fig:Braga1}, despite
yet imperfect squeezing of the contour lines in the right top panel. 
Nevertheless, the spectral function becomes 
well peaked at $T = 150$~MeV, see right bottom panel of figure~\ref{fig:Braga1}.

\begin{figure}[tb]
\center
\resizebox{0.99\columnwidth}{!}{%
\includegraphics{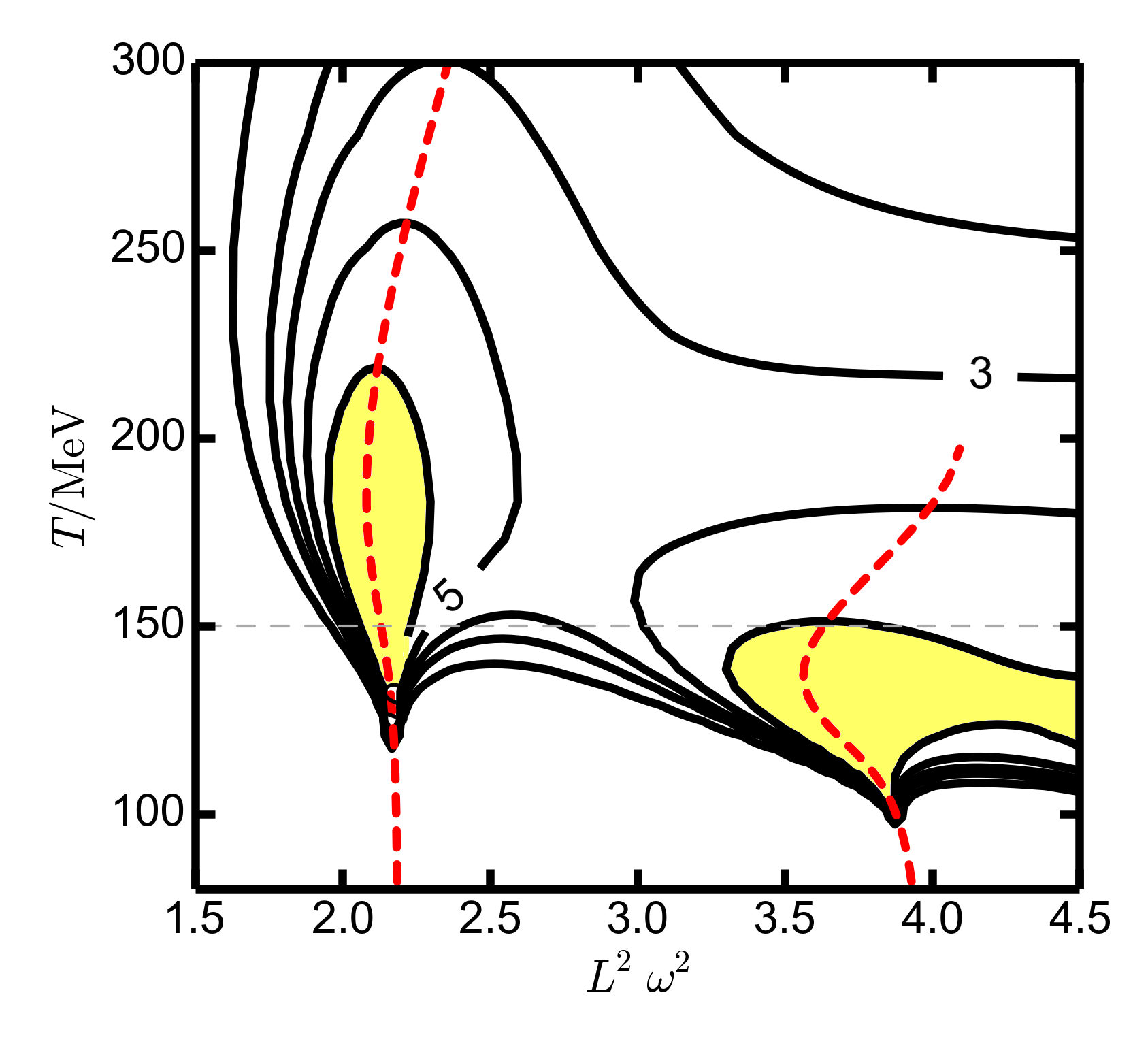}
\includegraphics{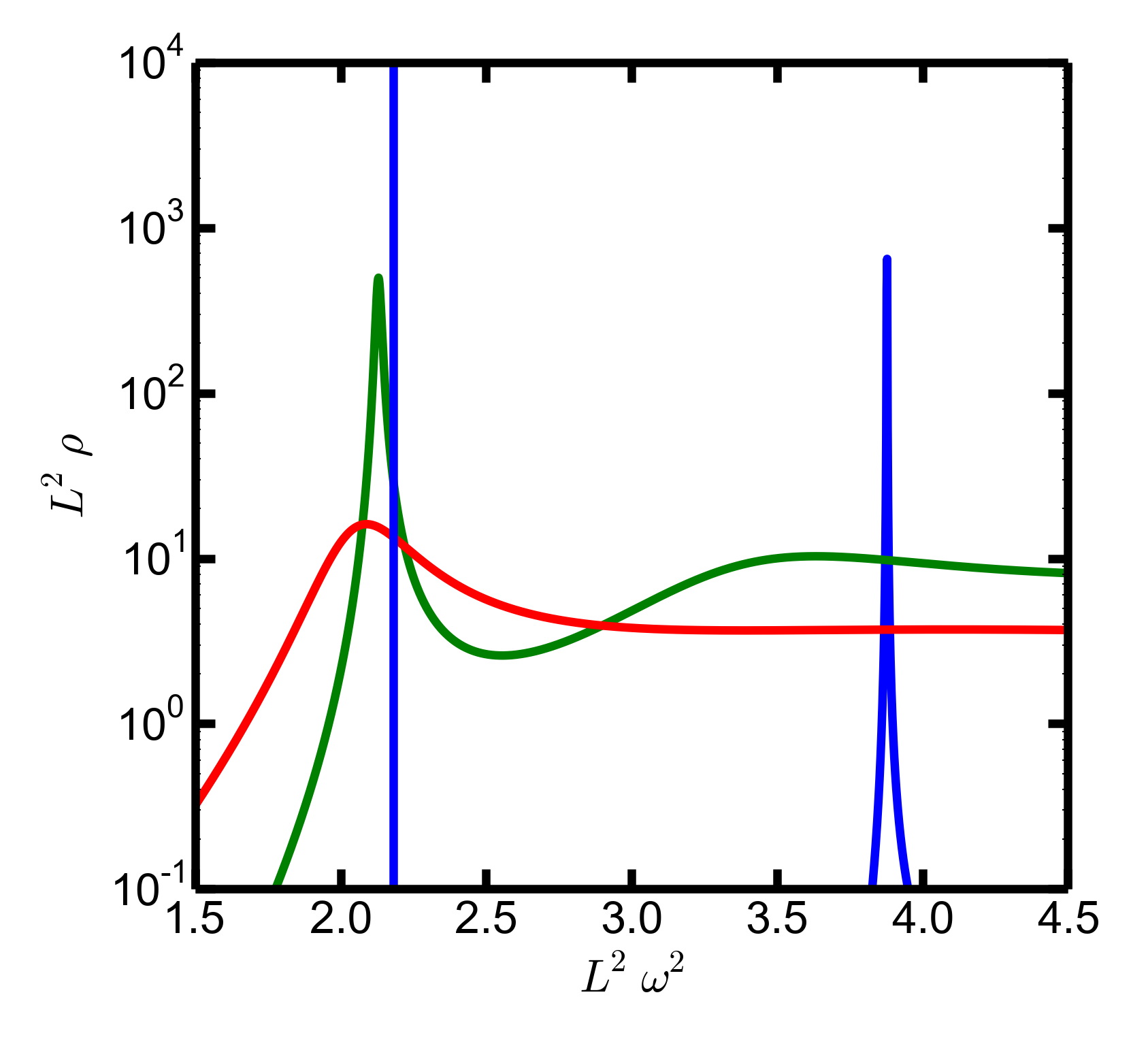}
\put(-743,141){{\Huge -----------------------------------------}}
\put(-545,255){{\huge $J/\psi$}}
\put(-200,255){{\huge $J/\psi$}}
}
\caption{Charmonium formation in the potential (\ref{eq:Braga_U0})
for the parameter set $(\sqrt{\Gamma}, k, M) = (0.55, 1.2, 2.2)$~GeV 
favored in \cite{Braga:2018zlu} 
but here combined with the QCD-related background.
Left panel: Contour plot of the charmonium spectral function $L^2 \rho$.
The dashed red curves are for the first two peak positions
($T_{melt}^{{\rm g.s.}} = 464$~MeV).
The dashed horizontal line indicates $T = 150$~MeV.
Right panel: Spectral functions $L^2 \rho$ at temperatures of 
100 (blue), 150 (green), 200 (red)  MeV.
\label{fig:Braga4}
}
\end{figure}

Finally, we exhibit in figure~\ref{fig:Braga4} the contour plot of 
the charmonium spectral function (left panel) and the spectral function
at selected temperatures (right panel) for the parameter set
$(\sqrt{\Gamma}, k, M) = (0.55, 1.2, 2.2)$~GeV favored in \cite{Braga:2018zlu}.
These parameters, albeit with noticeably deviations to the PDG values
of $m_{0, 1}$, realize the charmonium formation as transition of the
spectral function to a narrow, quasi-particle state at temperatures 
slightly below $T_c$. While the squeezing of the contour lines
near $T_c$ in the left panel of figure~\ref{fig:Braga4} is apparently not so
pronounced as in the case of bottomonium (see right top panel
in figure~\ref{fig:A2}), the spectral function displays a sharp peak at $T_c$,
see right panel of figure~\ref{fig:Braga4}. Insofar, it is justified to
speak on charmonium formation at $T_c$ for the given parameter set.
We emphasize the QCD-related background employed here, in contrast to the
schematic background in \cite{Braga:2018zlu}.

\begin{figure}
\centering
\resizebox{0.55\columnwidth}{!}{%
\includegraphics{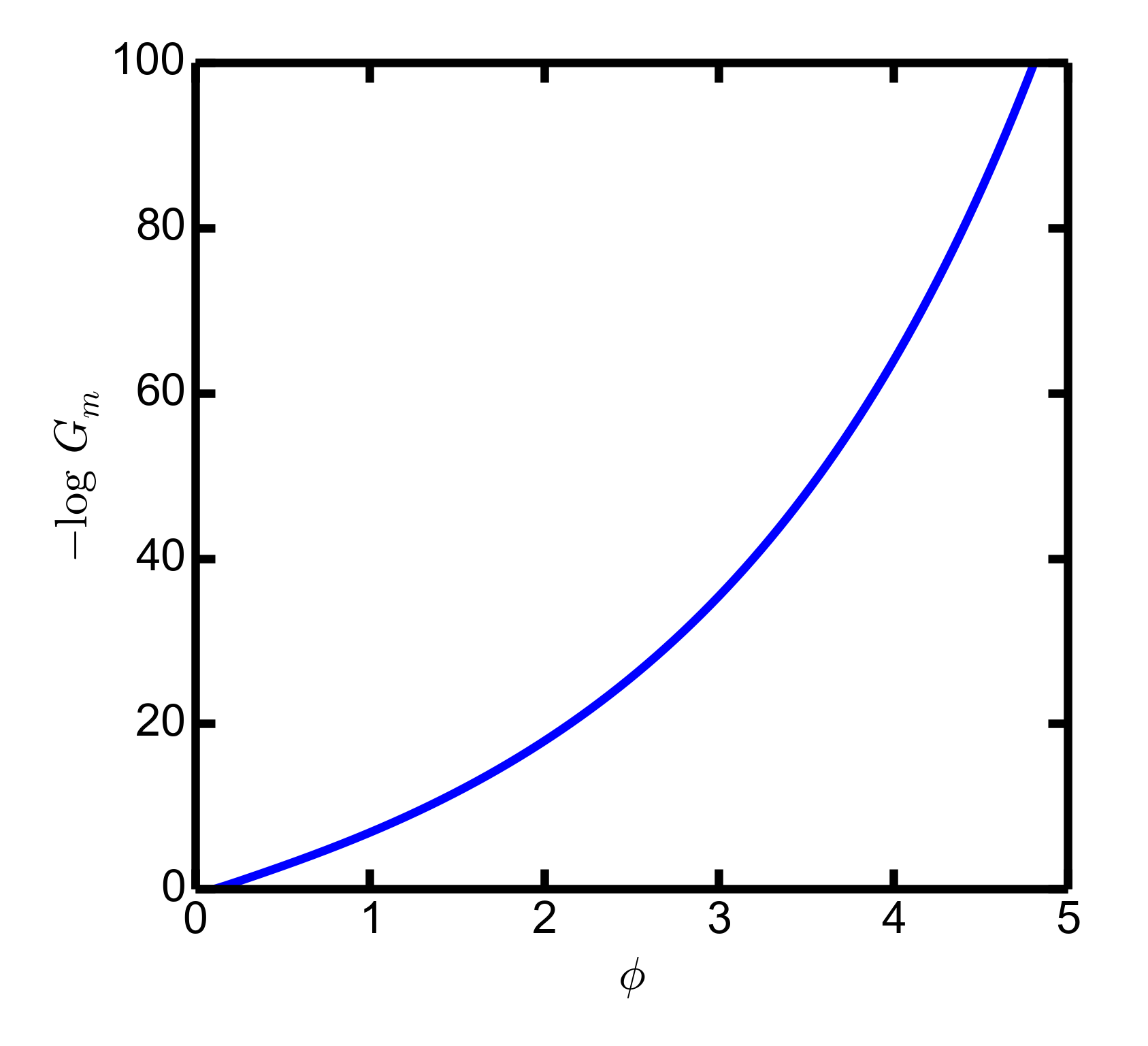}
}
\caption{The quantity $- \log G_m (\phi)$ calculated from
eqs.~(\ref{eq:8}, \ref{eq:9}) with $A_0'$, $\phi_0'$ deduced from
the thermodynamics in appendix \ref{App:B}
and the potential (\ref{eq:Braga_U0}) with
$(\sqrt{\Gamma}, k, M) = (1.0, 0.825, 3.656)$~GeV
yielding PDG values of charmonium masses $m_{0, 1}$.
$G_m(\phi)$ is supposed to be independent of temperature.
\label{fig:Braga5}
}
\end{figure}

To complete the systematic related to charmonium 
we exhibit in figure~\ref{fig:Braga5} the quantity
$- \log G_m$ as a function of $\phi$. 
Note the huge variation of $G_m (\phi)$. 
In general, $G_m (\phi)$ depends sensitively on the
parameters in $U_0$ and is tightly related to the 
background. 
 
An analog study of the $\Upsilon$ formation is hampered by some
uncomfortable structures of $U_0 (z; M, k, \Gamma)$. References
\cite{Braga:2015lck,Braga:2016wkm,Braga:2017bml}
advocate parameters which avoid such obstacles, however,
result in a value of $m_0^2 (\Upsilon(1S))$ 
being only one half of the PDG value.
We therefore do not perform an analysis of the potential ansatz~(\ref{eq:Braga_U0}) in the QCD-related background
since the two-parameter potential (\ref{eq:10}) was already
shown to accomplish successfully bottomonium formation at $T_c$.

\section{Summary}\label{sect:summary}

In summary we introduce a modification of the holographic vector meson action
for quarkonia such to join 
(i) the QCD$_{2+1}$(phys) thermodynamics, described 
dynamically consistently by a dilaton and the metric coefficients in AdS + BH, 
with (ii) realistic quarkonia masses at zero temperature.
Both pillars, thermodynamics and quarkonium mass spectra, 
are anchored in QCD as a common footing.
The formal holographic construction is
based on an effective dilaton $\phi_m = \phi - \log G_m$, where $\phi$ is
solely tight to the light-quark--gluon thermodynamics background, 
while the flavor dependent quantity $G_m$ is determined by
a combination of $\phi$ and the adopted Schr\"odinger equivalent potential
$U_0$ at zero temperature. $U_0$ encodes the flavor (or quark mass)
dependence and can be chosen with much sophistication to accommodate
many quarkonia  properties. We explore here the systematic of a two-parameter
model to demonstrate features of our scheme, 
where the thermodynamic background at $T > 0$ and meson spectra at $T = 0$
serve as QCD-based input to analyze the quarkonia formation at $T > 0$. 
We test a scenario where quarkonium formation is considered 
as an adiabatic process, 
i.e.\ a sequence of equilibrium states, 
and characterized by the shrinking of the respective
spectral functions towards narrow quasi-particle states,
in qualitative agreement with lattice QCD studies \cite{Larsen:2019zqv}.
Realistic values of $\Upsilon(1S, 2S)$ masses allow in fact the formation
temperature $T_{form}$
of $\Upsilon (1S)$ nearby $T_c$ in line with the claim of 
\cite{Braun-Munzinger:2018hat,Andronic:2017pug} 
that hadrons form themselves at temperatures 
$T_c \approx T_{fo}  \approx 155$~MeV.
Insofar, the mystery ``why $T_{fo} \approx T_c$?" could be resolved
by a dynamical process within such a scenario:
Hadronization is the transit of broad to narrow spectral functions
within a few-MeV temperature interval at $T_c$.  

While quite promising, the
proposed scenario is hampered by three issues, at least.
First, the finding of $T_{form} \approx T_c$ looks somewhat accidental
and is not locked explicitly to a certain microscopic process; in addition, there
is a slight tension due to the tendency of $T_{form} < T_c$ when deploying
the exact PDG value of the $\Upsilon (2S)$ mass
together with the $\Upsilon (1S)$ PDG value.  
Second, the formation of the $\Upsilon (2S)$ quasi-particle occurs
at $T_{form} ^{\Upsilon(2S)} < T_c$ due to the sequential formation,
which however could be an artifact of the two-parameter model of $U_0$.
Third, the envisaged scenario fails quantitatively
for $J/\psi$ since $T_{form}^{J/\psi} < T_c$ for the two-parameter model. 
It happens, however, that an improved, three-parameter model $U_0$
overcomes such problems to some extent, i.e.\
charmonium formation at $T_c$ is accomplished.
An ideal choice of $U_0$ should deliver the quarkonia mass spectra
(and other properties as well) and quarkonia formation as
rapid shrinking of the spectral functions in a narrow temperature
interval at $T_c$, including the excited states. 

Formally, hadronization of heavy-flavor probe quarkonia 
is determined by the potential $U_0$,
which governs the crucial function $G_m$, thus partially decoupling
it from the holographic background.

The here proposed bottom-up scenario of quarkonia formation solely accommodates
properties of vector $c \bar c$ and $b \bar b$ states in the holographic bulk
vector field ${\cal A}$. This is in contrast to microscopic studies,
e.g.\ in 
\cite{Strickland:2019ukl,Yao:2018sgn,Yao:2017fuc,Hoelck:2016tqf,Du:2019tjf,Du:2017qkv}, 
where the heavy-quark interaction with
constituents of the ambient medium is dealt with in detail.
Also primordial contributions and early off-equilibrium yields 
as well as corresponding feedings are not accounted for.
An important (yet) missing issue of the proposed scenario is a direct relation
to observables in relativistic heavy-ion collisions. All this calls for further investigations.

\appendix

\section{Specific features of the holographic gravity-dilaton background 
adjusted to QCD thermodynamics} \label{App:B}

The QCD$_{2+1}$(phys) equation of state obeys certain features.
Among them are the minimum of the sound velocity at 
$T \approx 145$~MeV, $v_s^2 (145~{\rm MeV}) \approx 0.15$,
and the maximum of the interaction measure at $T \approx 200$~MeV,
$(e - 3p)/T^4\vert_{200~{\rm MeV}} \approx 4$ 
\cite{Borsanyi:2013bia,Bazavov:2014pvz}.
The contributions of charm and bottom quarks are negligible at
$T \lesssim 200$~MeV \cite{Borsanyi:2016ksw}. The quoted temperature values
bracket the pseudo-critical temperature $T_c = (156 \pm 1.5)$~MeV
which is determined by a peak of the chiral susceptibility \cite{Bazavov:2018mes}.
We focus here on the local minimum of the sound velocity and its mapping
onto the gravity-dilaton background. 
 
Deforming the AdS metric by putting a black hole with horizon at $z_H$
yields the metric for the infinitesimal line elements squared (\ref{eq:3}) where $f(z, z_H) \vert_{z = z_H} = 0$ is a simple zero. Identifying the
Hawking temperature $T(z_H) = - \partial_z f(z, z_H) \vert_{z = z_H} /4 \pi$
with the temperature of the system at bulk boundary $z \to 0$
and the attributed Bekenstein-Hawking entropy density 
$s(z_H) = \frac{2 \pi}{\kappa} 
\exp\{ \frac32 A(z, z_H)\vert_{z=z_H} \}$,
one describes holographically the thermodynamics.
$f = 1$ at $T = 0$ refers to the vacuum. 

The gravity-dilaton background is determined by the action in the Einstein frame 
\begin{equation} \label{eq:S_grav}
S = \frac{1}{2 \kappa} \int \dd^4 x \, \dd z \sqrt{g_5}
\left[R - \frac12 (\partial_z \phi)^2 - V(\phi) \right] ,
\end{equation} 
where $R$ stands for the curvature invariant and $\kappa = 8 \pi G_5$. 
(For our purposes, the numerical values of $\kappa$ and $G_5$ as well as $k_V$
in (\ref{eq:1}) are irrelevant.)
The field equations and equation of motion for the metric coefficients and the dilaton
follow from (\ref{eq:S_grav}) as
\begin{eqnarray}
A'' &=& \frac12 A'^2 - \frac13 \phi'^2 , \label{A:3}\\
f'' &=& - \frac32  A' f' ,  \label{A:4}\\
\phi'' &=& -\left( \frac23 A' +\frac{f'}{f} \right) \phi' + 
\frac{1}{f} e^A \partial_\phi V \label{A:5}
\end{eqnarray}
to be solved with boundary conditions
$A(z \to 0) \to -2 \log (z/L)$, 
$\phi(0) = 0$, $\phi'(0) = 0$,
$f(0) = 1$, $f(z_H) =0$; the prime means differentation w.r.t.\ $z$.
The dilaton potential $V(\phi)$ is the central quantity \cite{Zollner:2018uep}.
Imposing certain conditions one can describe the QCD-relevant cross-over 
(instead of phase transitions of first or second order or a Hawking-Page transition).
A necessary condition for a cross-over is (i) $\partial_\phi V / V $, as a function of $\phi$, has a local maximum
and (ii) $\partial_\phi V / V < \sqrt{2/3}$ 
(for refinements, cf.~\cite{Zollner:2018uep}).

The three-parameter ansatz
\begin{equation} \label{eq.A2}
- L^2 V = 12 \cosh(\gamma \phi) + \phi_2 \phi^2 + \phi_4 \phi^4
\end{equation}
is sufficient for a satisfactory description of the lattice QCD$_{2+1}$(phys) data
\cite{Borsanyi:2013bia,Bazavov:2014pvz}\footnote{More 
parameters are required for a perfect match of
the various thermodynamic state variables as a function of the
temperature within the full data range, cf.\ figure~1 in \cite{Knaute:2017opk}
and further references therein.}
by coefficients $(\gamma, \phi_2, \phi_4) = (0.568, -1.92, -0.04)$
together with $L^{-1} = 1.99$~GeV,
see figure~5-left in \cite{Zollner:2020cxb}. 
In fact, the above mentioned conditions are met:
maximum of $\partial_\phi V / V = 0.58$ at
$\phi = 1.84$.

\begin{figure}[tb]
\center
\resizebox{0.99\columnwidth}{!}{%
\includegraphics{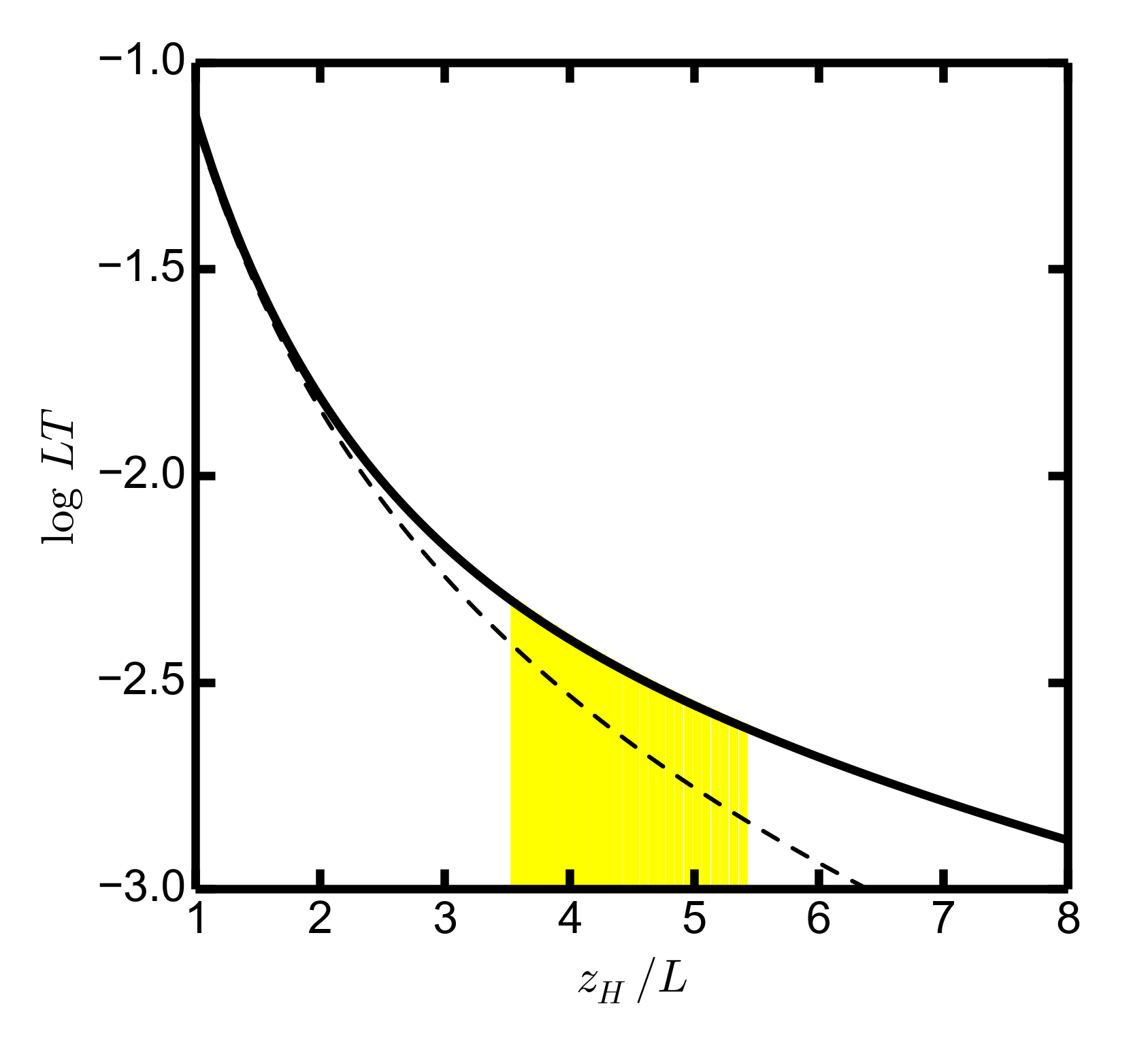}
\includegraphics{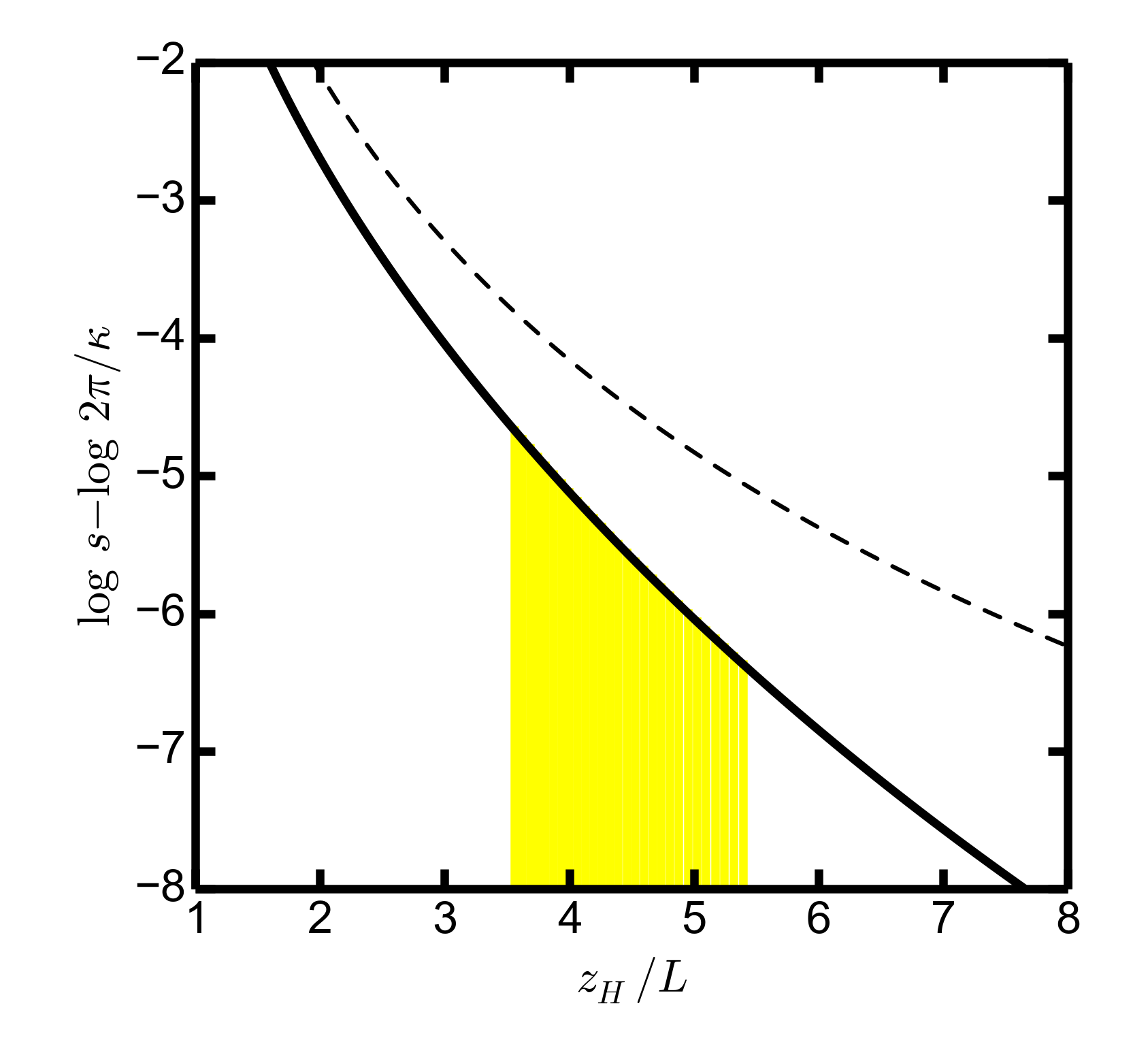}
}
\caption{The logarithms of temperature, $\log LT$ (left panel), 
and entropy density,
$\log s - \log 2 \pi /\kappa = \frac32 A(z_H, z_H)$ (right panel), 
as a function of $z_H / L$.
Dashed curves are for $LT = 1 / (\pi z_H)$ (left) and
$\frac32 A = - 3 \log z_H / L$ (right).
The colored regions are for $z_H = [z_H^{v_s^2}, z_H^{\hat I}]$,
which are determined by $T_c^{v_s^2} = 145$~MeV (position of the
minimum sound velocity) and $T_c^{\hat I} = 200$~MeV 
(position of the maximum of interaction measure $\hat I = (e - 3 p) /T^4$) 
according to \cite{Bazavov:2014pvz}.
\label{B:1}
}
\end{figure}

\begin{figure}
\centering
\resizebox{0.50\columnwidth}{!}{%
\includegraphics{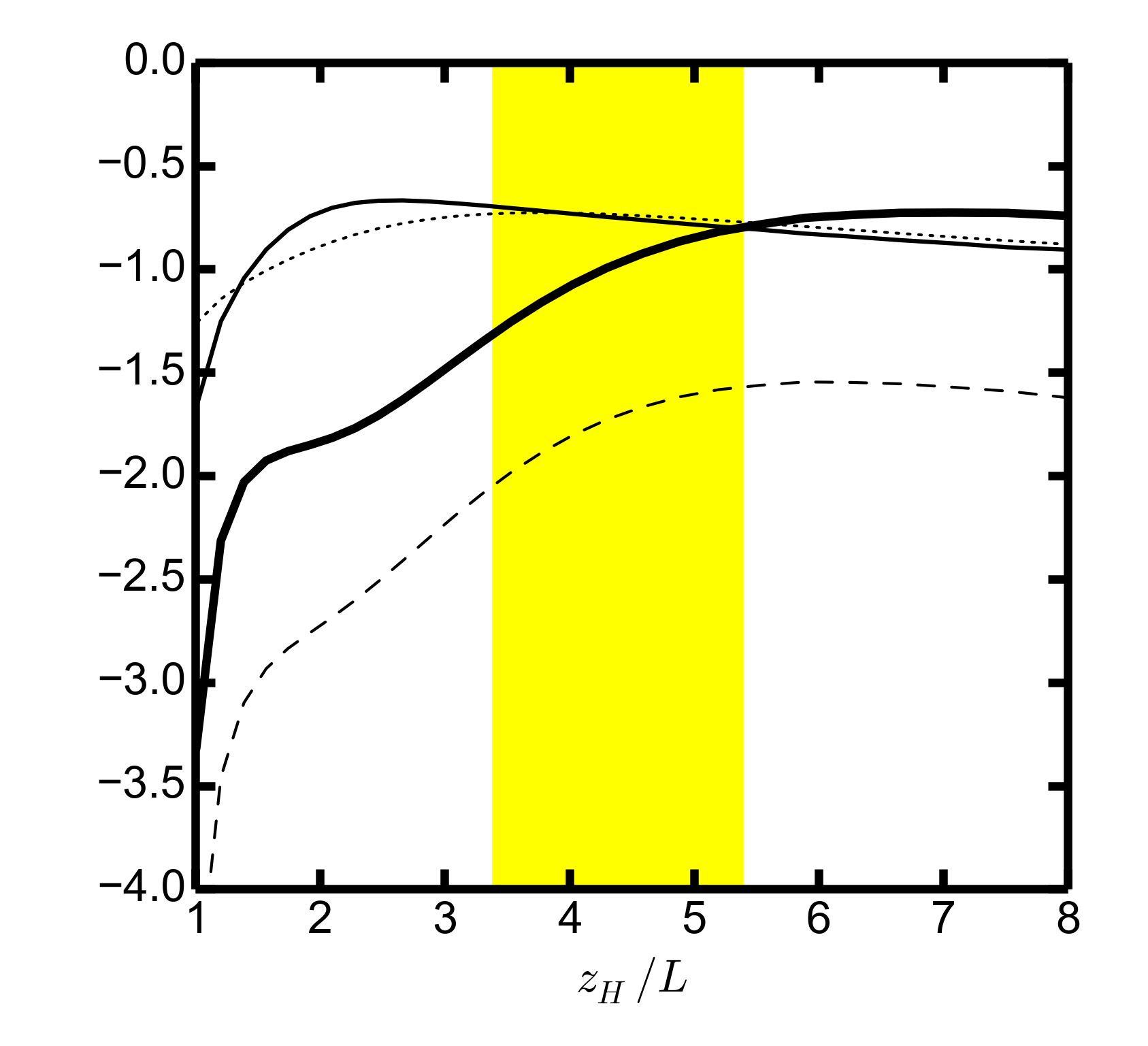}
\put(-310,111){{\Large $\partial_{z_H}^2 T /  \partial_{z_H} T$}}
\put(-310,263){{\Large $\partial_{z_H} T /  T$}}
\put(-210,250){{\Large $\partial_{z_H}^2 T /  \partial_{z_H} T
- \partial_{z_H} T /  T$}}
\put(-322,315){{\Large $\partial_{z_H}^2 A /  \partial_{z_H} A$}}
}
\caption{The terms entering (\ref{min_vs2})
multiplied by $z_H$:
$\partial_{z_H}^2 T /  \partial_{z_H} T$ - dashed curve, 
$\partial_{z_H} T /  T$ - dotted curve, 
$\partial_{z_H}^2 T /  \partial_{z_H} T
- \partial_{z_H} T /  T$ - fat solid curve, 
$\partial_{z_H}^2 A /  \partial_{z_H} A$ - thin solid curve.
The yellow region is as in figure~\ref{B:1}.
\label{B:2}
}
\end{figure}

In general, the sound velocity squared, $v_s^2 = \frac{d \log T}{d \log s}$,
acquires a local minimum if $s(T)$, or $s (T) /T ^4$, has an inflection
point. Surprisingly, neither $T(z_H)$ nor $s(z_H)$ display such a feature.
Instead, both $T(z_H)$ and $s(z_H)$ are monotonous functions of $z_H$,
see figure~\ref{B:1}. That means, the minimum of the sound velocity
is caused by a subtle interplay of derivatives of $T(z_H)$ and $s(z_H)$.
Displaying the sound velocity squared by
$v_s^2 (z_H) = \partial_{z_H} \log T / \partial_{z_H} \log s$,
the local minimum is determined by
\begin{equation} \label{min_vs2}
\partial_{z_H}^2 T /  \partial_{z_H} T 
- \partial_{z_H} T /  T 
- \partial_{z_H}^2 A (z = z_H) /  \partial_{z_H} A (z = z_H)= 0.
\end{equation}
These individual terms are exhibited in figure~\ref{B:2}.
It turns out that the actually chosen parameters facilitate
the minimum of sound velocity
at the crossing of the fat solid and thin solid curves at
$z_H /L  = 5.17$, corresponding to $T = 152$~MeV, i.e.\
nearby $T_c$ and thus $T_{fo}$. 

\begin{figure}[h]
\center
\resizebox{0.99\columnwidth}{!}{%
\includegraphics{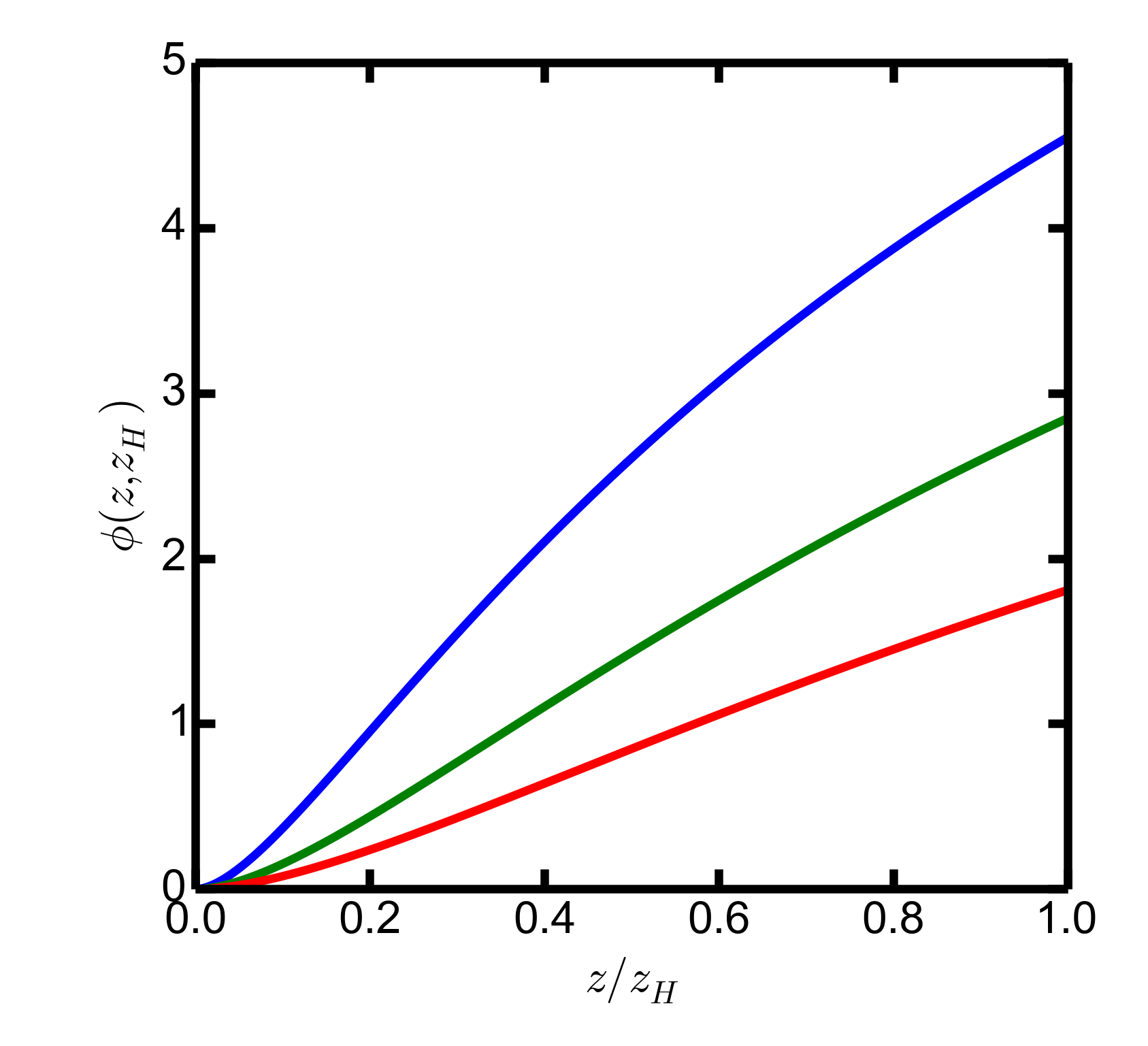}
\includegraphics{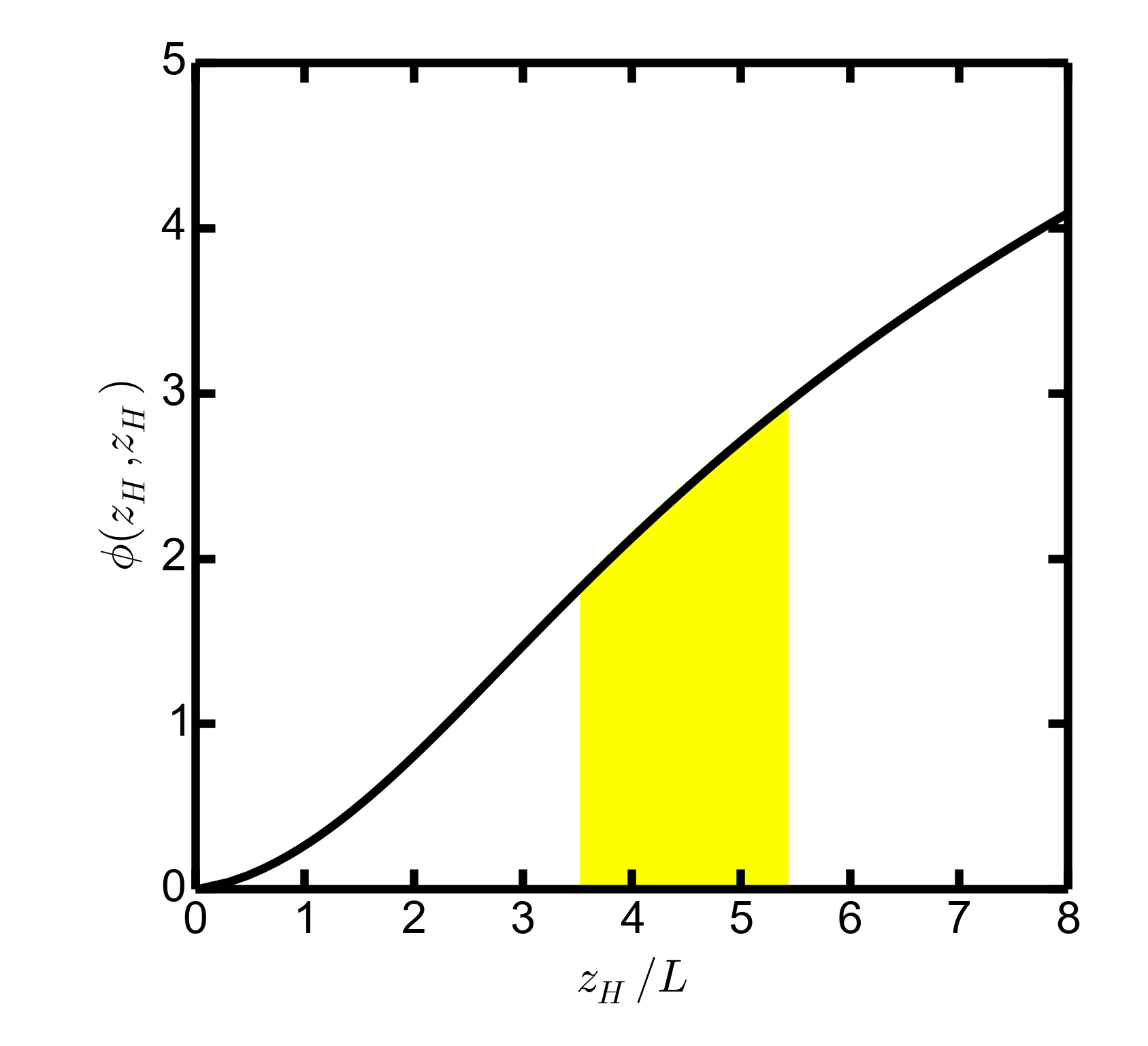}
\put(-655,270){{\huge {\color{blue} $T =100$ MeV}}}
\put(-550,210){{\huge {\color{green} $150$ MeV}}}
\put(-550,90){{\huge {\color{red} $200$ MeV}}}
}
\caption{The dilaton profile $\phi (z, z_H)$ 
as a function of $z/z_H$ (left panel, 
for $z_H / L = 9.28$, 5.22 and 3.52 corresponding to $T = 100$ (blue), 150 (green), 200 (red) MeV
with inflection points at $z/z_H =  0.20$, 0.33, 0.44) 
and $\phi(z, z_H)\vert_{z = z_H}$ as a function of $z_H$ (right panel,
the inflection point is at $z_H/L = 2.91$).
\label{B:3}
}
\end{figure}

In contrast to $T(z_H)$ and $A(z, z_H)\vert_{z = z_H}$,
the dilaton profile $\phi (z, z_H)$ has a marked imprint of the QCD specifics: it exhibits inflection points
in both $z$ direction and $z_H$ direction, see figure~\ref{B:3}.
This is a remarkable property which makes the use of the QCD-related gravity-dilaton background distinct
in comparison with schematic ans\"atze, which additionally miss the consistent interrelations of the quantities $A$, $f$ and $\phi$
via field equations. Note that the dilaton enters explicitly the quarkonium action (\ref{eq:1}),
thus leaving directly its imprints related to quarkonium formation.

\acknowledgments
The authors gratefully acknowledge the collaboration with J.~Knaute
and thank
M.~Ammon, P.~Braun-Munzinger,
M.~Kaminski, K.~Redlich and G.~R\"opke for useful discussions.
The work is supported in part by the European Union’s Horizon 2020 research
and innovation program STRONG-2020 under grant agreement No 824093. 


\end{document}